\documentclass[useAMS,usenatbib]{mn2e}

\usepackage{graphics}
\usepackage{graphicx}
\input{epsf}
\begin{document}

\def\vs{\bf Varun}
\def\ss{\bf Sergei:}
\def\bc{\begin{center}}
\def\ec{\end{center}}
\def\b{\begin{equation}}
\def\e{\end{equation}}
\def\ber{\begin{eqnarray}}
\def\eer{\end{eqnarray}}
\def\l{\left}
\def\r{\right}
\def\eg{{\it e.g.~}}
\def \ie {{\em i.e.~~}}
\def \lleq {\lower0.9ex\hbox{ $\buildrel < \over \sim$} ~}
\def \ggeq {\lower0.9ex\hbox{ $\buildrel > \over \sim$} ~}
\def \dlt {$\delta$}
\def \L {$\Lambda$}
\def \T {$\tau$}
\def \l {$\lambda$}
\def \hm1 {$h^{-1}$}
\def \mf {Minkowski functionals}

\title[Paper 2]
{Morphology of the supercluster-void network in \L CDM cosmology}
\author[S.F. Shandarin, J.V. Sheth, V. Sahni]
{Sergei F.Shandarin$^{1,3}$, 
Jatush V.Sheth$^{2,4}$ 
and Varun Sahni$^{2,5}$  \\
  $^{1}$ Department of Physics and Astronomy, University of Kansas,KS
  66045, USA \\
  $^{2}$ Inter University Centre for Astronomy $\&$ Astrophysics, Pune, 
India \\
$^{3}$ sergei@ku.edu\\
$^{4}$ jvs@iucaa.ernet.in\\
$^{5}$ varun@iucaa.ernet.in
}

  \maketitle
\begin{abstract}
We report here the {\em first} systematic study of the supercluster-void
network in the $\Lambda$CDM concordance cosmology in which voids and superclusters are 
treated on an equal footing. Superclusters are defined as individual 
members of an over-dense 
excursion set and voids are defined as individual members of 
a complementary under-dense 
excursion set at the same density threshold.
We determine the geometric, topological and morphological properties of the 
cosmic web at a large set of density levels by computing Minkowski 
functionals for every
supercluster and void using SURFGEN \citep{sh-s-s-sh03}.
The properties of the largest (percolating) supercluster and the complementary
void are found to be very different from properties of individual superclusters
and voids. Individual superclusters totally occupy no more than about 5\% of
the total volume and contain no more than 20\% of mass if the largest
supercluster is excluded. Likewise,
individual voids totally occupy no more than 14\%  of volume
and contain no more than 4\% of mass if the largest void is excluded.
Although superclusters are more massive and voids are more
voluminous the difference in maximum volumes is not greater than by an order
of magnitude. The genus value of individual superclusters can 
be $\sim 5$ while 
the genus of individual voids can reach $\sim 40$, implying significant
amount of substructure in superclusters and especially in voids.
One of our main results is that large voids, 
as defined through the density field (read dark matter
distribution) can be distinctly non-spherical.

\end{abstract}

\section{Introduction}
\label{sec:intro}
One of the great observational discoveries of recent times is the
realization that we live on a cosmic web which is embedded in an
accelerating Universe.  Although an accelerating universe was
originally established using high redshift type Ia supernovae
\citep{sn1,sn2}, the case for a `dark energy' dominated universe has
received independent support from observations of the cosmic microwave
background \citep{wmap} combined with studies of galaxy clustering in
the two degree field (2dF) galaxy redshift survey \citep{efs02}.
Although the nature of dark energy is still an open issue, a
cosmological constant appears to agree very well with all current
observations which indicate $\Omega_\Lambda \simeq 0.73$, $\Omega_m
\simeq 0.23$ and $\Omega_b \simeq 0.04$ (for reviews of dark energy
see \citet{ss00,sahni02}).
Fully three dimensional large scale galaxy catalogues reveal that the
cosmic web consists of an interpenetrating network of superclusters
and voids.  It therefore becomes important to understand and quantify
the geometrical and topological properties of large scale structure in
a \L CDM cosmology in a deep and integrated manner.

The main aim of the present study is to study the supercluster-void
network in \L CDM cosmology with emphasis on the sizes, shapes and
topologies of individual superclusters and voids.  We shall also study
the percolation properties of the full excursion set sampling the
entire density field and quantify our results in terms of Minkowski
functionals (hereafter, MFs).  A study such as the present one will
help us in comparing theory against observations.
It will also help us to distinguish the salient features of \L CDM
cosmology from sister cosmologies in which dark energy is a function
of time.  (These issues will be discussed in detail in companion
papers.)

It may be appropriate to mention that this is the first comprehensive
analysis of large scale structure geometry and morphology in which
over-dense (superclusters) and under-dense (voids) regions are treated
on a completely equal footing.  Earlier studies have emphasized either
over-densities (clusters, superclusters) or under-densities (voids) as
a result of which the methods used for the analysis of these two
complementary entities (superclusters/voids) often vary greatly in the
literature.  Thus over-dense regions have been studied using
correlation functions, minimal spanning trees, shape functions etc.,
whereas voids have been constructed from point processes using
elaborate boundary and volume filling techniques (see \eg 
\citet{sc95,masaar} and references therein).  Although the above
methods do provide us with some insights into supercluster-void
morphology, we shall follow an alternative route in this paper and
study the properties of the supercluster-void network using the
density field as our main starting point and fundamental physical
quantity. The reason for this is two fold, firstly, density fields can
be easily constructed from point data sets using, for example, cloud
in cell (CIC) techniques (this is true both for data from N-body
simulations, which we examine in this paper, as well as for galaxy
distributions in three dimensions). Secondly, an elaborate surface
modeling scheme, SURFGEN (short for `surface generator'), allows us to
determine the geometry, morphology and topology of excursion sets
defined on a density field in a very comprehensive manner
\citep{sh-s-s-sh03}. Applying SURFGEN to the density distribution in
the \L CDM model allows us to develop deep insights into the
distribution of large scale structure in this model, which can be
quantified using MFs and Shapefinders. Working with the density field
also permits us to determine the morphological properties of the {\em
  full excursion set} describing the supercluster-void network. More
detailed information is then gleaned at one particular threshold
(usually associated with percolation) at which shapes and sizes of individual
superclusters and voids yield rich information about properties of the
cosmic web to which we belong.

Concretely, we study the large scale structure of the universe by
considering the geometry and topology of isodensity surfaces
$\delta({\bf x})\equiv \delta \rho({\bf x})/\bar{\rho}=const$.  At a
given threshold $\delta_{\rm TH}$ regions having higher than threshold
density ($\delta > \delta_{\rm TH}$) will be called ``superclusters'',
while regions with $\delta < \delta_{\rm TH}$ will be called ``voids''.
Thus, we define superclusters and voids as over-dense and
under-dense connected regions bounded by one (or several) surfaces of
constant density.  This definition broadly corresponds to
other definitions of superclusters and voids used in the literature
but differs in details.  Apart from obvious differences with
superclusters and voids of galaxies in redshift space our approach
specifies neither a particular density threshold nor the shapes of the
structures. Despite these differences we call the objects of our study
superclusters and voids mostly because they are nonlinear structures
having sizes, volumes and masses roughly corresponding to 
superclusters and voids of galaxies.

We study superclusters and voids at a large number of
density thresholds and construct the isodensity surfaces at
every threshold to best accuracy.  In contrast to many studies (see \eg
\citet{blu_etal92,gol_vog03}), we do not `cook up' voids or
superclusters with predefined shapes but isolate individual objects
from the dark matter density field obtained in the N-body simulations
by building the isodensity surfaces.  Apart from uniformly smoothing
the density field with a Gaussian filter, we do not introduce any factors
that may affect the shapes or substructure in superclusters and voids
of the cosmic density field.  Filtering the density field may erase
some small-scale features but it certainly does not introduce any 
new structures. Thus, we
know beforehand that the real structure can be only richer and more
elaborate compared to what we study after smoothing. Filtering high
frequency modes is virtually implied in every physics study.  This
approach can be also viewed as an application of the standard
excursion set technique to non-Gaussian three-dimensional fields.

One of the goals of this work is a quantitative and objective testing
of some stereotypes routinely used in cosmology. Here are a few
examples: both N-body simulations and redshift surveys are
characterized by filamentary and sheet-like or pancake-like
structures; voids are quasi-spherical bubbles in the density and/or
galaxy distributions; and expand faster than the universe as whole;
voids occupy most of the volume in the universe. The first problem
arises when one tries to address these issues in the absence of
conventional definitions. Frequently the context of the origin and
usage of a particular assertion is either obscure or forgotten.
Proving that some of the above cliches cannot be true does not require
elaborate N-body simulations or sophisticated analysis.  For instance,
voids cannot {\em both} occupy most of the universe and expand faster
than the universe because this would require the expansion of voids in
{\em comoving space}. But this is impossible since the comoving volume
is conserved !  Therefore, the expansion of some voids must clearly be
at the expense of some of the others.  This is one of the trivial (but
important) conclusions arising from the adhesion model
\citep{gurbatov89,sh_z89,sss94}.  At the same time it is not necessary
for the interior of the squeezed voids to shrink.  The interior of the
void may continue to expand faster than the Hubble flow but the void
eventually vanishes because its boundaries move {\em inwards}
\citep{sss94,sss95a,sss95b}. So as the void collapses it is likely to
become increasingly non-spherical. Thus, a single expanding void model
while being reasonable for a crude estimate of substructure growth may
be completely misleading for estimating the sizes of a realistic
ensemble of voids.

Pancakes or sheet--like structures remain an unsolved controversy since
their theoretical prediction by Zel'dovich in 1970. On the one hand
there were numerous assertions that they had been detected in the
N-body simulations or redshift surveys most of which were based on 
visual impressions.  On the other, many believe that pancakes are not
clearly seen in any realistic N-body simulations and have not been
detected objectively.  It is worth noting that the simulation of the
structure in the hot dark matter model did not show the existence 
of pancakes \citep{kly_sh83} which was interpreted as the result of
weak singularities corresponding to pancakes compared to considerably 
stronger
singularities corresponding to filaments \citep{ashz82}. The coarse graining
unavoidable in N-body simulations erases pancakes while filaments
survive. This argument has been essentially repeated by \cite{bon_kof_pog96}
and elaborated for the probabilistic interpretation of the large-scale
structure. In this study of the \L CDM model we do not see pancakes
either.

The large--scale structure in the dark matter density field is
certainly different from the one observed in galaxy redshift surveys.
Galaxies are at best pointwise tracers of the parent continuous dark
matter density field. It is likely that they are biased tracers, and
that they do not display real physical structures but structures which are
strongly distorted by the mapping into redshift space. 
Thus, the observed superclusters and voids are not physical but only
apparent objects as the retarded motion of some planets is only
apparent but not real motion.  
At present there are only two methods of investigating real large--scale
objects: by reconstructing the real space density field  from  
peculiar velocities of galaxies and by investigating the density
fields in the models that are consistent with observations such as \L CDM.
  
  We consider this study of the dark matter density field in real
  space as a first step in a systematic study of the morphology of 
  large-scale structure.  It will be followed by studies of mock and
  real galaxy catalogues.  Understanding the morphology of the dark
  matter distribution in real space is an important component in
  understanding the physical processes determining the formation of
  galaxies and their motions as well as for building theoretical
  models of superclusters and voids.

Due to the fairly large smoothing scale adopted here, the over-density in
superclusters ranges from $\delta \sim 1$ to $\delta \sim 10$, which
makes them more extended ($ \ggeq $ few Mpc) and considerably less dense 
than galaxy clusters in (for instance) the Abell catalogue.

The rest of the paper is organized as follows.  In Section
\ref{sec:simulations} we briefly describe the density fields used in
this study. In Section \ref{sec:morph_param} we define and very
briefly discuss the morphological measures which we apply to study the cosmic web. Sec.
\ref{sec:perc_topol} provides the relation of the filling
factors used in this study with the probability density function.
Sec. 5 discusses the morphological properties of individual superclusters 
and voids. Sec. 6 describes substructure in superclusters and voids.
A discussion of some properties of Minkowski functionals is contained in
Sec. 7. Section 8 summarizes our main results.

\section{Dark Matter Distribution in Virgo Simulations}
\label{sec:simulations}
We use dark matter distributions in a flat model with $\Omega_0$ =
0.3, $\Omega_{\Lambda}$ = 0.7, $h=0.7$ (\L CDM). The initial spectrum
was taken in the form suggested by \cite{efs_bon_whi92} with the shape
parameter $\Gamma = 0.21$.

The amplitude ($\sigma_8=0.9$) of the power spectrum in the model is
set so as to reproduce the observed abundance of rich galaxy clusters
at the present epoch.  A detailed discussion of the cosmological
parameters and simulations can be found in \citet{jenkcdm}.

The density fields generated by the VIRGO simulations have been
studied in several papers (see e.g. \cite{springtop,scf99}); this is
the first systematic study of global and local morphology when both
over-dense and under-dense regions are measured by the same method.

SURFGEN operates on three-dimensional pixelized maps. Therefore we
first generate the density field from the distribution of dark matter
particles.  This process was described in detail in
\citet{sh-s-s-sh03}; here we present a brief summary.  The data
consist of $256^3$ particles in a box of size 239.5 \hm1 Mpc. We fit a
$128^3$ grid to the box. Thus, the size of each cell is 1.875 \hm1
Mpc.  Here we follow the smoothing technique used by \citet{springtop}
which they adopted for their preliminary topological analysis of the
Virgo simulations.  In the first, we apply a Cloud in Cell (CIC)
technique to construct a density field on the grid.  Next we smooth
this field with a Gaussian kernel which offers us an extra smoothing
length-scale.  We study the field smoothed with $L_s = 5$ \hm1 Mpc but
also present the global MFs for the field smoothed with $10$ \hm1 Mpc
for comparison.  The scale of $L_s = 5$ \hm1 Mpc is a fiducial
smoothing scale in many studies of both density fields in N-body
simulations and galaxy fields from redshift surveys; see for example, 
\cite{gro_gel00}.  The Gaussian kernel for smoothing that we adopt
here is
\b W(r) = {1 \over (\pi)^{3\over2} L_s^3} exp{\left(-{r^2 \over
      L_s^2}\right)}.  \e
Since the kernel is isotropic, it is likely to diminish the true
extent of anisotropy in filaments and pancakes.  This effect could be
minimized by considering anisotropic kernels and/or smoothing
techniques based on the wavelet transform.  An even more ambitious
approach is to reconstruct density fields using Delaunay tesselations
using a technique reported by \cite{rien}. Density fields reconstructed
in this manner appear to preserve anisotropic features and may
therefore have some advantage over conventional `cloud-in-cell'
techniques followed by an isotropic smoothing \citep{sw00}. Thus, as
far as one of the goals is to utilize the geometry of the patterns to
discriminate between the models, such smoothing schemes should prove
more powerful. We hope to return to these issues in further
publications.

We scan the density fields at 99 values of the density threshold, 
all equispaced in the filling factor $FF_C$ defined as
\b
\label{eq:FF-def} 
FF_{\rm C}(\delta_{\rm TH}) = { 1 \over V_{tot}}\int \Theta(\delta-\delta_{\rm TH}) d^3x, 
\e 
where $\Theta(x)$ is the Heaviside Theta function and $\delta=(\rho
-\bar{\rho})/\bar{\rho}$ is the density contrast.  The supercluster
filling factor $FF_{\rm C}$ measures the volume-fraction in regions
which satisfy the `supercluster' criterion $\delta_{\rm sc} \geq
\delta_{\rm TH}$. In the
following, we use $FF_{\rm C}$ along with the void filling factor
$FF_{\rm V} \equiv 1-FF_{\rm C}$ as a parameter to label the density
contours. The relation between $FF_{\rm C}$ and density contrast
threshold can be seen in Fig. \ref{fig:cpf}.

At each level of the density field (labeled by $FF_{\rm C}$ or
$FF_{\rm V}$), we construct a catalogue of superclusters (over-dense
regions) and voids (under-dense regions) based on a grid realization
of the Friends of Friends (FOF) algorithm.

Next we (i) run the SURFGEN code on each of these superclusters/voids
to model surfaces for each of them and (ii) determine the Minkowski
Functionals (MFs) for all superclusters/voids at the given threshold
(these are referred to in the literature as partial MFs). Global MFs
are partial MFs summed over all superclusters.  Thus, at each level of
the density, we first compute the partial MFs and then the global MFs.
\section{Morphological Parameters}
\label{sec:morph_param}
Since the 1970s, theory, N-body simulations, and, most importantly,
galaxy redshift surveys have strongly suggested that the components of
the large-scale structure can be roughly placed in three classes:
compact quasi-spherical or slightly elliptical structures like Abell
clusters, long filaments like the famous `bridge' connecting the Coma
cluster and A1367, (\cite{gre_tho78}) and voids.  There have also been
claims that pancake-like concentrations of galaxies have been observed
\cite{fairall,masaar}.  The voids have often been claimed to have
quasi-spherical or slightly elliptical shapes. Most of these claims
have been based on visual impressions. In particular, statistics used
in studies of voids often assumed that voids are spherical or close to
spherical thereby precluding any other possibility.

In this work we study the geometry and topology of the regions bounded
by the isodensity surfaces and therefore make no prior assumptions about
the shapes of superclusters and voids.  It is worth noting that some
regions may have more than one boundary surface and possess nontrivial
topology of the boundaries.  The complete characterization of an
arbitrarily complex region in three dimensions obviously cannot be
achieved if only a few numbers are used. At best one can try to design
some basic characteristics that serve a particular purpose. Our
purpose is to provide basic measures suitable for quantification of
typical components of the large-scale structure: superclusters and voids.

Four Minkowski functionals are effective non-parametric descriptors of
the morphological properties of surfaces in three dimensions
\citep{meckwag94,matsub03,sh-s-s-sh03}. They are
\begin{itemize}
\item {\it Volume} $V$ enclosed by the surface, $S$,
\item {\it Area} $A$ of the surface,
\item {\it Integrated mean curvature} $C$ of the surface, 
\b
\label{eq:curv}
C = \frac{1}{2}\oint_S{\left({1\over R_1} + {1\over R_2}\right)da}, 
\e
where $R_1$ and $R_2$ are the principal radii of curvature at a given
point on the surface.
\item {\it the Euler characteristic}
\b
\label{eq:euler} \chi = \frac{1}{2\pi}\oint_S{\left({1\over
      R_1R_2}\right)da}.  
\e
\end{itemize}
Although for our purpose the Euler characteristic is a more convenient
quantity than genus, $G$ (see Appendix for a discussion) the latter
has been used more often in cosmology. The genus is uniquely related
to the Euler characteristic $G = 1 - \chi/2$ and thus carries exactly
the same information. We measure the above parameters for every region
in both over-dense and under-dense excursion sets at 99 density
thresholds equispaced in the filling factor from $FF=0.01$ to
$FF=0.99$.

As demonstrated in \citet{sss98,sss98b} particular ratios of Minkowski
functionals called ``Shapefinders'' provide us with a set of
non-parametric measures of sizes and shapes of objects.  Therefore, in
addition to determining MFs we shall also derive the Shapefinders, $T$
(Thickness), $B$ (Breadth) and $L$ (Length) defined as follows:
\b \label{eq:TBL}
T = {3V\over A}, \ \  B = {A\over C}, \ \ L = {C\over 4\pi}.
\e 
The three Shapefinders describing an individual region bounded by one
or several isolated surfaces of constant density have dimensions of
length and provide us with an estimate of the regions `extension': $T$ is the
shortest and thus describes the characteristic thickness of the region or
object, $L$ is typically the longest and characterizes the length of
the object; $B$ is intermediate and can be associated with the breadth
of the object.  This simple interpretation is obviously relevant only
for fairly simple shapes.  The choice of the coefficients in eq.
\ref{eq:TBL} results in a sphere having all three sizes equal to its
radius $T=B=L=R$. A triaxial ellipsoid has values of $T$, $B$ and $L$
close but not equal to the lengths of its three principal semi-axes:
shortest, intermediate and the longest respectively.  It is worth
noting that $T$, $B$ and $L$ are only the estimates of three basic
sizes (semi-axes) of an object which work quite well on such objects
as a triaxial ellipsoid and torus \citep{sss98,sss98b,sh-s-s-sh03}
but no three numbers can describe an arbitrary, complex
three-dimensional shape.

An indicator of `shape' is provided by a pair of dimensionless
Shapefinder statistics
\b 
\label{eq:shapefinder} 
P = {B-T\over B+T};~~~ F = {L-B\over L+B}, 
\e 
where $P$ and $F$ are measures of Planarity and Filamentarity
respectively ($P, F \leq 1$).  A sphere has $P = F = 0$, an ideal
filament has $P = 0, F = 1$ while $P = 1, F = 0$ for an ideal pancake.
Other interesting shapes include `ribbons' for which $P \sim F \sim
1$.  When combined with the genus measure, the triplet $\lbrace P,F,G
\rbrace$ provides an example of {\em shape-space} which incorporates
information about topology as well as morphology of superclusters and
voids.\footnote{Non-geometrical shape-statistics based on mass moments
  etc. can give misleading results when applied to large scale
  structure, as demonstrated in \cite{sss98b}.}
\section{Percolation and global topology}
\label{sec:perc_topol}
\subsection{Filling factors and one-point function}
\begin{figure}
\centering
\centerline{
  \includegraphics[width=3.5in]{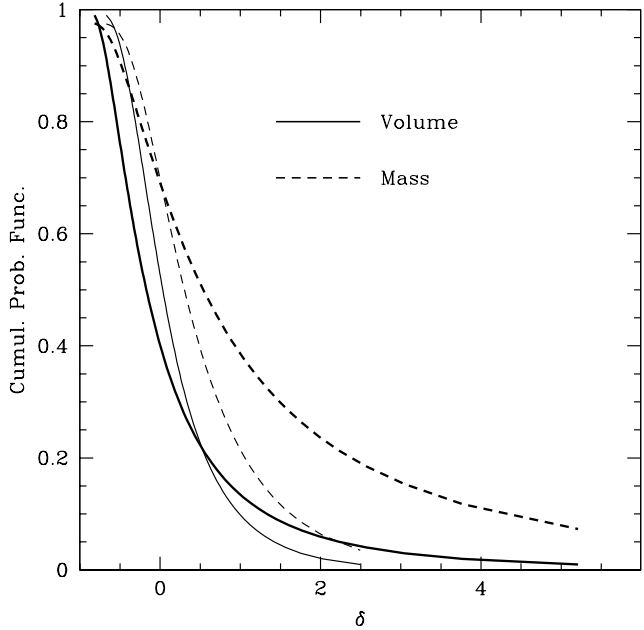} }
\caption{Cumulative probability functions of the density contrast
  in the Virgo simulations of the \L CDM model smoothed with $L_s$=5
  \hm1 and 10 \hm1 Mpc are shown by the thick and thin solid lines
  respectively.  The fractions of mass in the excursion sets are
  shown by the dashed line.  The cumulative probability function
  equals the filling factor $FF_C$.}
\label{fig:cpf}
\end{figure}

A characteristic of the density field which is both simple and useful
is the one-point probability density function, $p(\delta)$, where
$\delta \equiv (\rho - \bar{\rho})/\bar{\rho}$.  In this study we use
the integral
\begin{equation}
FF_C(\delta)= P(\delta) = \int\limits_{\delta}^{\infty} p(\delta') d\delta'
\label{eq:ffc}
\end{equation} 
known as the cumulative probability function (cpf) as a major
quantity parameterizing the excursion sets. 
It measures the fraction of volume in the excursion set,
$\delta > \delta_{\rm TH}$. In order to compare supercluster and void
parameters we also use the under-density filling factor
\begin{equation}
FF_V(\delta)= 1-FF_C(\delta) = \int\limits_{-1}^{\delta} p(\delta') d\delta'.
\label{eq:ffv}
\end{equation} 
The filling factor $FF_C=FF_C(\delta)$ is shown in Fig. \ref{fig:cpf}
(thick solid line)
for \L CDM smoothed on the length scale of $L_s$=5 \hm1 Mpc.  In
addition, the thick dashed line shows the fraction of mass in the excursion
set for the same smoothing scale. 
From Fig. \ref{fig:cpf} we find that the density field is
`nonlinear' ($\delta >1$) in a relatively small fraction (\lleq 15\%)
of the total volume.  However, the difference between these two curves is
the first clear demonstration of nonlinearity of the density field.
Convergence to the Gaussian distribution with the growth of the
smoothing scale is demonstrated by the two thin lines
corresponding to the smoothing scale of $L_s$=10 \hm1 Mpc.
\subsection{Percolation}
\begin{figure}
\begin{minipage}[t]{.99\linewidth}
  \centering\includegraphics[width=4.cm]{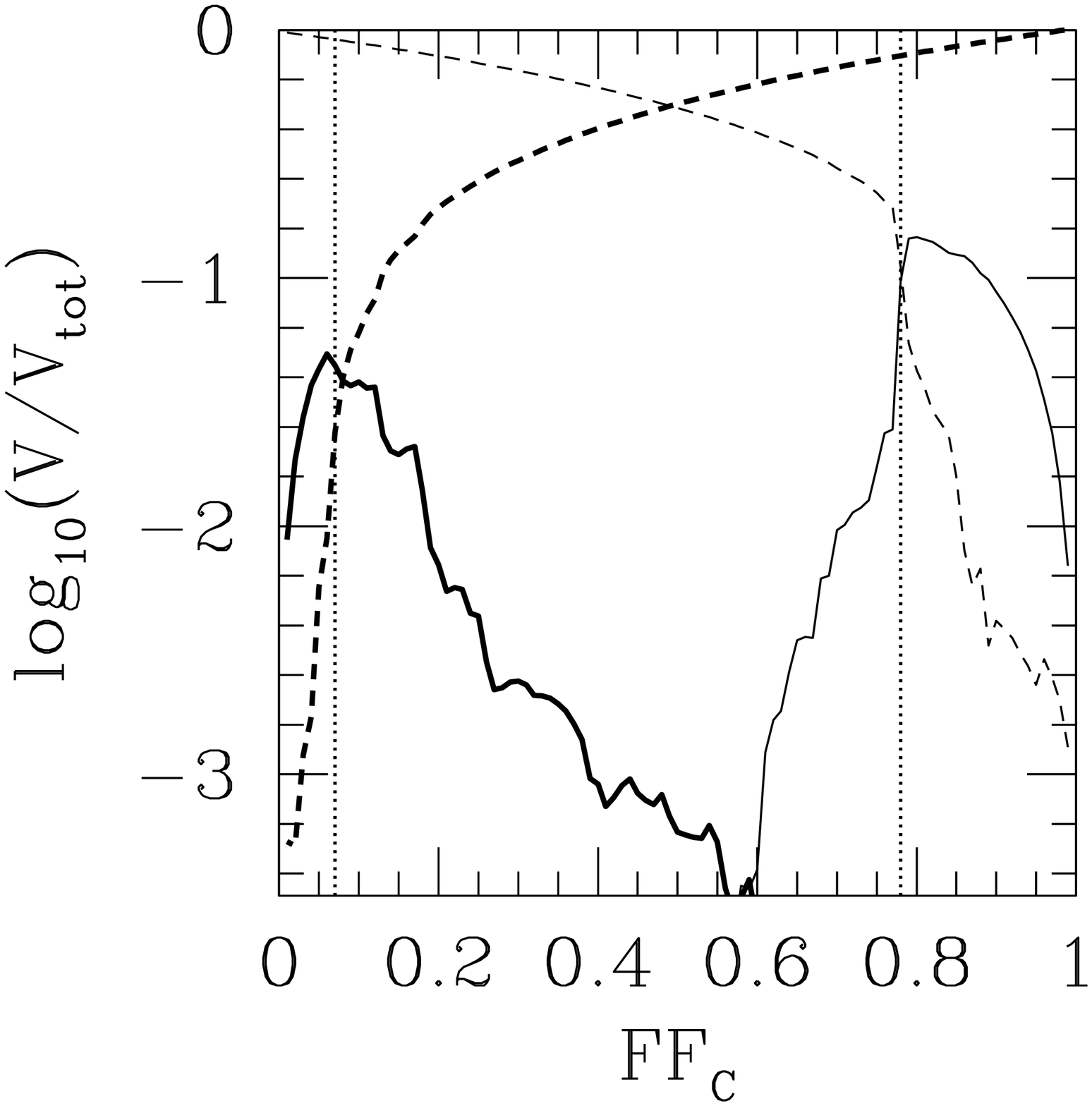} 
  \includegraphics[width=4.cm]{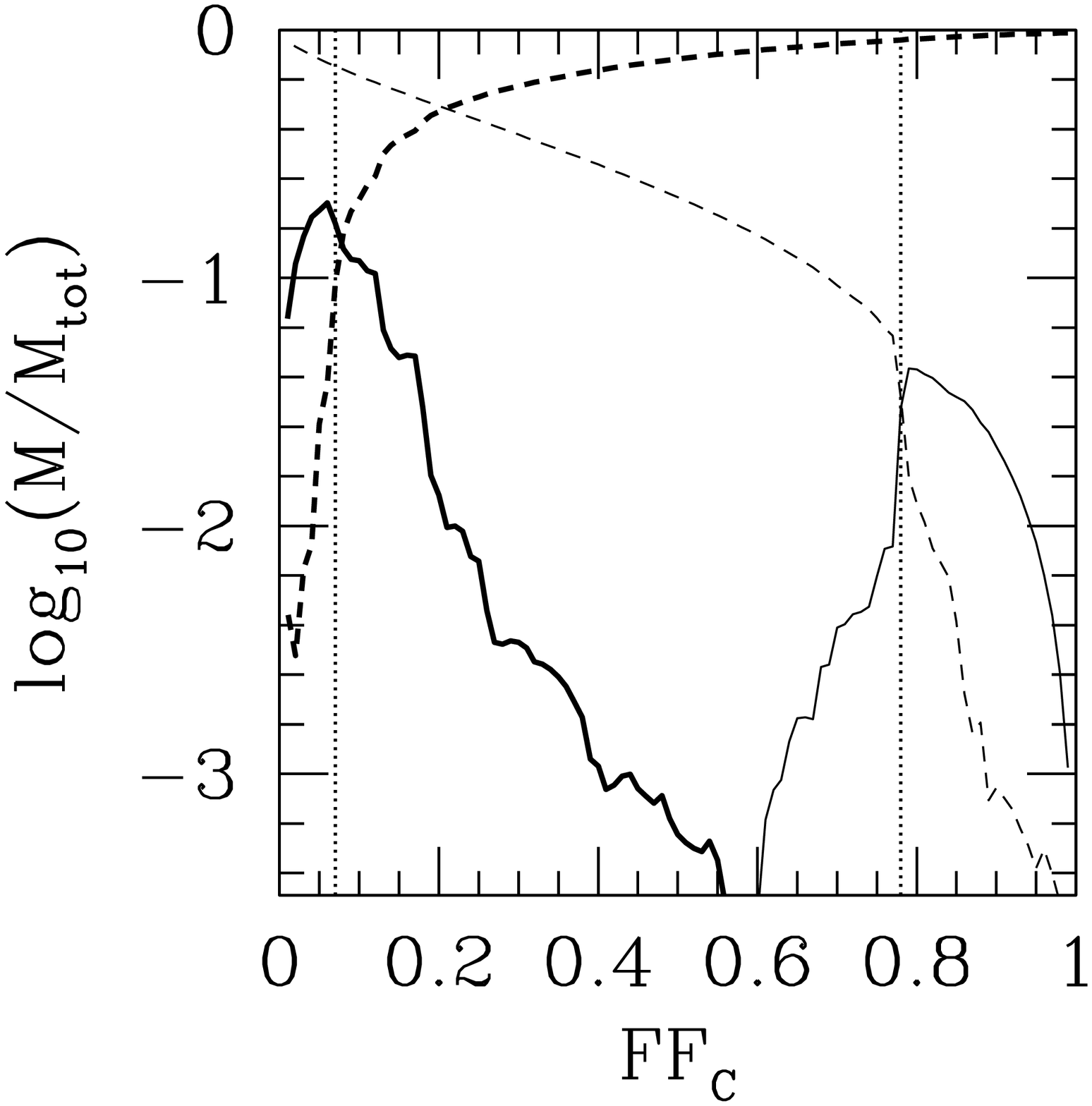}
\end{minipage}\hfill
\caption{Left panel: the fractions of the total volume occupied by 
  the largest supercluster (thick dashed line), all superclusters but
  the largest one (thick solid line), largest void (thin dashed line),
  and all voids but the largest one (thin solid line) are shown for
  the density field in the \L CDM model smoothed with $L_s$=5 \hm1 Mpc
  as a function of the filling factor, $FF_C$.  Right panel: the
  y-axis shows the fraction of mass in the components shown in the
  left panel. Vertical dotted lines show the percolation threshold for
  superclusters ($FF_C \approx 0.07$) and voids ($FF_C \approx 0.78$;
  note: $FF_V=1-FF_C \approx 0.22$).  }
\label{fig:isol_vm}
\end{figure}

Understanding percolation is essential for understanding the morphology of
the supercluster-void network. Percolation is important because the
properties of superclusters and voids radically change at the
percolation transitions (see Fig. \ref{fig:isol_vm}). The left panel
shows: (i) the fractional volumes in the largest supercluster and void
(dashed lines) and (ii) the total volume in all superclusters and voids
after the largest object has been removed from the sample (solid lines).  
The right
panel shows mass fractions in all four components.  At relatively
high thresholds $\delta_C \ggeq 1.8$ corresponding to small filling
factors $FF_C \lleq 0.07$ the largest supercluster has insignificant
volume and mass compared to the total volume or mass contained in the
over-dense excursion set, $\delta > \delta_{\rm TH}$.  During the
percolation transition at $FF_C \approx 0.07$ corresponding to
$\delta_C \approx 1.8$, both volume and  mass in the largest
supercluster rapidly grow, overtaking the volume and mass in the entire
excursion set, and completely dominating the entire sample from 
this point onwards.
The largest void behaves in a qualitatively similar manner if plotted
versus $FF_V$. At $FF_V \lleq 0.22,\ \delta_V \lleq -0.5$ its volume
is small compared to the volume of the under-dense excursion set,
$\delta < \delta_{\rm TH}$, but at the percolation transition $FF_V
\approx 0.22,\ \delta_V \approx -0.5$, it takes over and from then 
on remains the
dominant structure in the under-dense excursion set.  Since $FF_C
\equiv 1-FF_V$ the void percolation transition takes place at $FF_C
\approx 0.78$ as shown in Fig. \ref{fig:isol_vm}.

Two obvious conclusions can be drawn from the above discussion.  First, at
percolation the object having the largest volume becomes very
different from all remaining objects, therefore it must be studied separately.
Second, individual objects -- both superclusters and voids -- must be
studied in the corresponding phase {\em before percolation occurs} in the
corresponding phase. Both superclusters and voids reach
their largest sizes, volumes and masses just before percolation sets in.

Figure \ref{fig:isol_vm} also shows that at $FF_C > 0.5$ superclusters
dominate by volume while at $FF_C < 0.5$ voids dominate.  In the range
$0.2 \lleq FF_C \lleq 0.7$ corresponding to $1.6 \ggeq \delta \ggeq
-0.43$, no more than 10\% of volume is occupied by non-percolating
objects, while the remaining more than 90\% of entire volume is
occupied by just two largest objects: percolating supercluster and the
percolating void. In the range between the two percolation
thresholds $FF_C \approx 0.07$ and $FF_C \approx 0.78$, both largest
objects percolate and therefore the density field has a sponge like
structure.  The interval of sponge like structure in a Gaussian
field is between $ FF_C \approx 0.16$ and $FF_C \approx 0.84$.
Therefore nonlinear gravitational evolution has shifted it toward
smaller $FF_C$ and increased its range by a little bit from 0.68 (=0.84$-$0.16)
to 0.71 (=0.78$-$0.07).

It is worth stressing that the shift and length of the sponge like
interval are determined by the percolation thresholds which in general
are two independent parameters.  As a result of these shifts the
interval of the so called ``meat-ball'' topology (when no supercluster
percolates) has reduced compared to the Gaussian case from $FF_C
\approx 0.16$ to $FF_C \approx 0.07$ and the interval of the
``bubble'' topology (when no void percolates) is increased from $FF_V
\approx 0.16$ to $FF_V \approx 0.22$.  All the numbers obviously
depend on the adopted smoothing scale but the sign and type of change
must be universal for the $\Lambda$CDM model.

For both superclusters and voids 
$FF=FF_{\rm max}+FF_{\rm ind}$, where $FF_{\rm ind}$
stands for the fractional volume occupied by all objects excluding
the largest one, and $FF_{\rm max}$ is the fractional volume in the 
largest object.
Further since, $FF_{\rm max}/FF_{\rm ind} \gg 1$ in the
the most part of the range between two  percolation thresholds
for both superclusters and voids, it is not surprising that
$FF_{\rm max}^{C} =FF_{\rm max}^{V}$ almost exactly at
$FF_C=FF_V=0.5$. A similar result was found in the models with 
power law initial spectra \citep{ys96}.

Percolation is characterized by many features, the most conspicuous 
being the rapid merger of disjoint parts of the excursion set into one
connected structure spanning the entire volume.  Merging of
superclusters occurs when the density threshold is reduced whereas
merging of voids correspond to the growth of the threshold. Spanning
of the largest supercluster or void throughout the whole volume
results in connection of the opposite faces of the cubic volume by this
structure which explains the term {\em percolation}.  Although in
principle the percolation transition can be determined by checking if
the opposite faces of the cube are connected, in practice it is more
robust to identify percolation using other properties of the excursion
set \citep{kly87,kly_sh93}.
\begin{figure}
\centering
\centerline{
  \includegraphics[width=3.in]{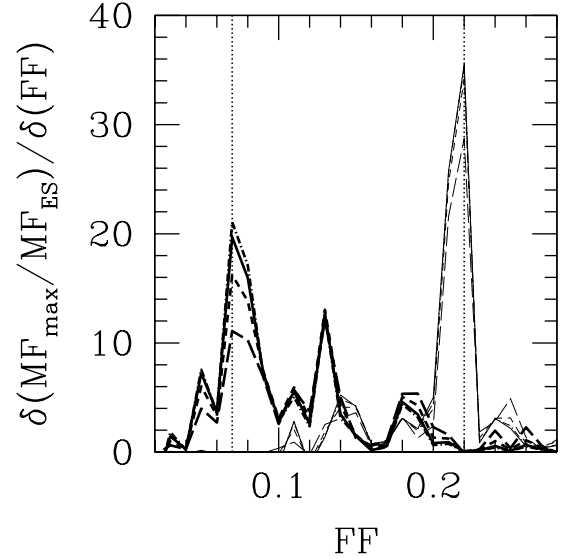} }
\caption{Estimates of the percolation thresholds.
  The rate of growth $\delta m^{(i)}/\delta (FF)$ 
 for the four estimators listed in eq. \ref{eq:lcs_def} is shown as a
  function of $FF$. Thick lines show results for superclusters.
  All four curves consistently peak at $FF=FF_C =
  0.07$.  Thin lines show similar quantities for voids with a distinct
  peak at $FF=FF_V = 0.22$. Solid, short dashed, long dashed, and
  dot-dashed lines show the volume, area, curvature, and mass
  estimators respectively. Vertical dotted lines mark the percolation
  thresholds.}
\label{fig:perc_estim}
\end{figure}
As the estimators of the percolation threshold we use the following 
four ratios 
\b 
m^{(i)} = \frac{MF^{(i)}_{\rm max}}{MF^{(i)}_{ES}},\ \ i=1,2,3,4, 
\label{eq:lcs_def}
\e
where $MF^{(i)}$ is one of four quantities: volume, area, integrated
mean curvature, or mass 
($MF^{(1)}=V,\ MF^{(2)}=A,\ MF^{(3)}=C, \ MF^{(4)}=M $). 
The subscript ``max'' is
self-explanatory, and ``$ES$'' stands for the entire excursion set. At
percolation these ratios grow extremely rapidly from very small values to
unity. The maximum rate of growth $\delta m^{(i)}/\delta (FF)$ can
be used as a reliable estimator of the percolation threshold as
shown in Fig.  \ref{fig:perc_estim}.   All four parameters detect the
percolation transitions at $FF_C \approx 0.07$ for superclusters 
and at $FF_V \approx 0.22$  for voids. These transitions
correspond to the density contrast $\delta_C \approx 1.8$ for
superclusters and $\delta_V \approx -0.5$ for voids.
\begin{figure}
\centering
\centerline{
  \includegraphics[width=3.3in]{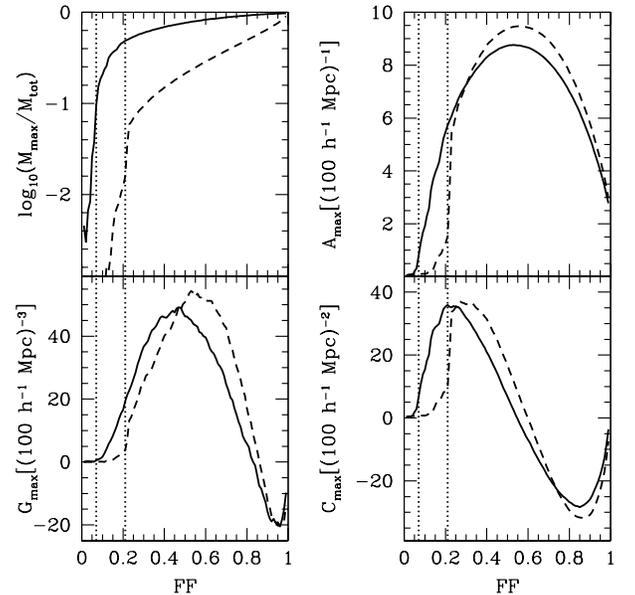} }
\caption{The mass fraction and Minkowski functionals of the largest 
  (by volume) supercluster and void in the \L CDM model smoothed with
  the L$_s$=5 h$^{-1}$ Mpc window as a function of corresponding
  filling factor ($FF = FF_C$ for the largest supercluster; $FF = FF_V$ for
  the largest void).  Solid and dashed lines show the parameters of
  the largest supercluster and void respectively. Vertical dotted
  lines mark the percolation transitions.}
\label{fig:perc_mf}
\end{figure}

Figure \ref{fig:perc_mf} compares the largest supercluster with
the largest void.  Three \mf~ and the fraction of the mass in the largest
objects are shown as a function of the corresponding filling factor.
The difference of two curves in every panel is a significant
indication of non-Gaussianity of the density field.  All Gaussian
curves must coincide in every panel and show maximum growth at
$FF^{SC}_C = FF^V_V \approx 0.16$.  The nonlinear gravitational
evolution in the \L CDM model shifts the percolation transition in
the over-dense excursion set toward smaller filling factors ($FF_C:\ 0.16 
\rightarrow 0.07$) and in 
the under-dense excursion set toward larger filling factors
($FF_V:\ 0.16 \rightarrow 0.22$). 
Again the particular numbers depend on the
choice of the smoothing scale but the sign of the effects is
independent of the smoothing scale. At smaller smoothing scales 
as well as for any adaptive smoothing having a better resolution
in high density regions the supercluster percolation threshold
must decrease ($FF_C < 0.07$) and void percolation threshold must 
increase ($FF_V > 0.22$).
On the other hand, increasing the smoothing scale would result in a
continuous reduction of differences between superclusters and voids
and their ultimate convergence to Gaussian
curves in every panel (not shown).

It is worth noting that the percolation transition in voids is more
conspicuous than that in superclusters. The transition is particularly
clearly marked by $A_{\rm max}$ and $C_{\rm max}$ curves.  All curves look
differently after percolation as well.  However, in order to 
precisely evaluate
the significance level of these differences one needs to analyze more
than one realization and/or have larger simulation volume. 
\subsection{Global topology}
It is interesting to compare the percolation and genus statistics.
Both were suggested as tests for assessing the connectedness of 
the large-scale structure.  In a series of papers Zel'dovich and Shandarin
\citep{z82,sh83,shz83} raised the question of topology of large-scale
structure and suggested percolation statistic as a discriminator
between models.  The percolation test was first applied to a redshift catalog
(compiled by J. Huchra) by \citet{zes82} and then \citet{ein-etal84}
who found that the connectivity between galaxies in this catalog was
significantly stronger than for a Poisson distribution. In contrast, a
non-dynamical computer  model having approximately correct
correlation functions up to the fourth order \citep{son-p78} 
showed significantly weaker connectivity than observed. 
Thus, percolation was able to detect connectedness in the galaxy distribution.
It was also demonstrated that three lowest order correlation functions  
(two-, three- and four-point functions) are not sufficient to detect
the connectedness in the galaxy distribution. 
\begin{figure}
\begin{minipage}[t]{.99\linewidth}
\centering
  \includegraphics[width=4cm]{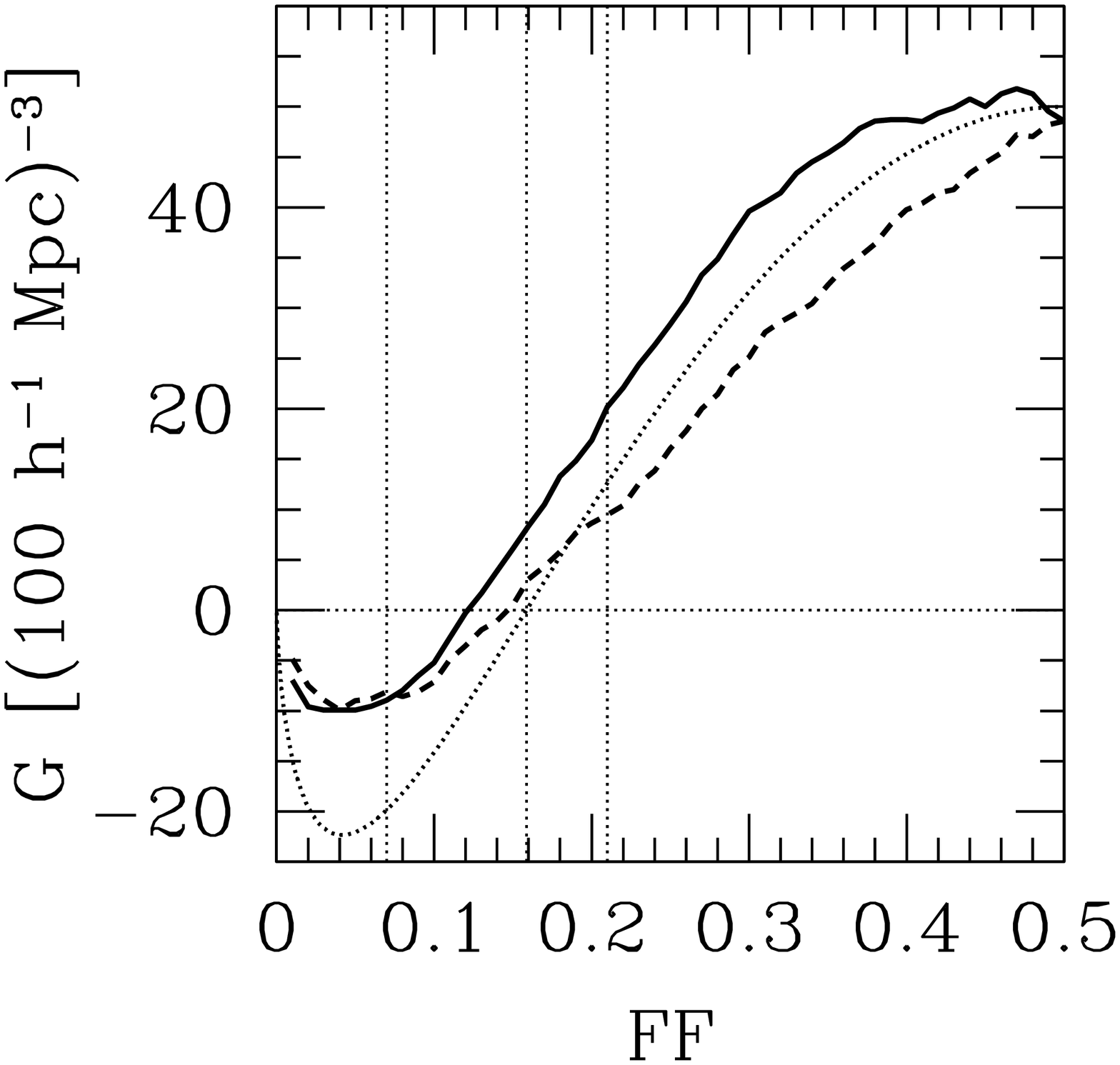}
  \includegraphics[width=4cm]{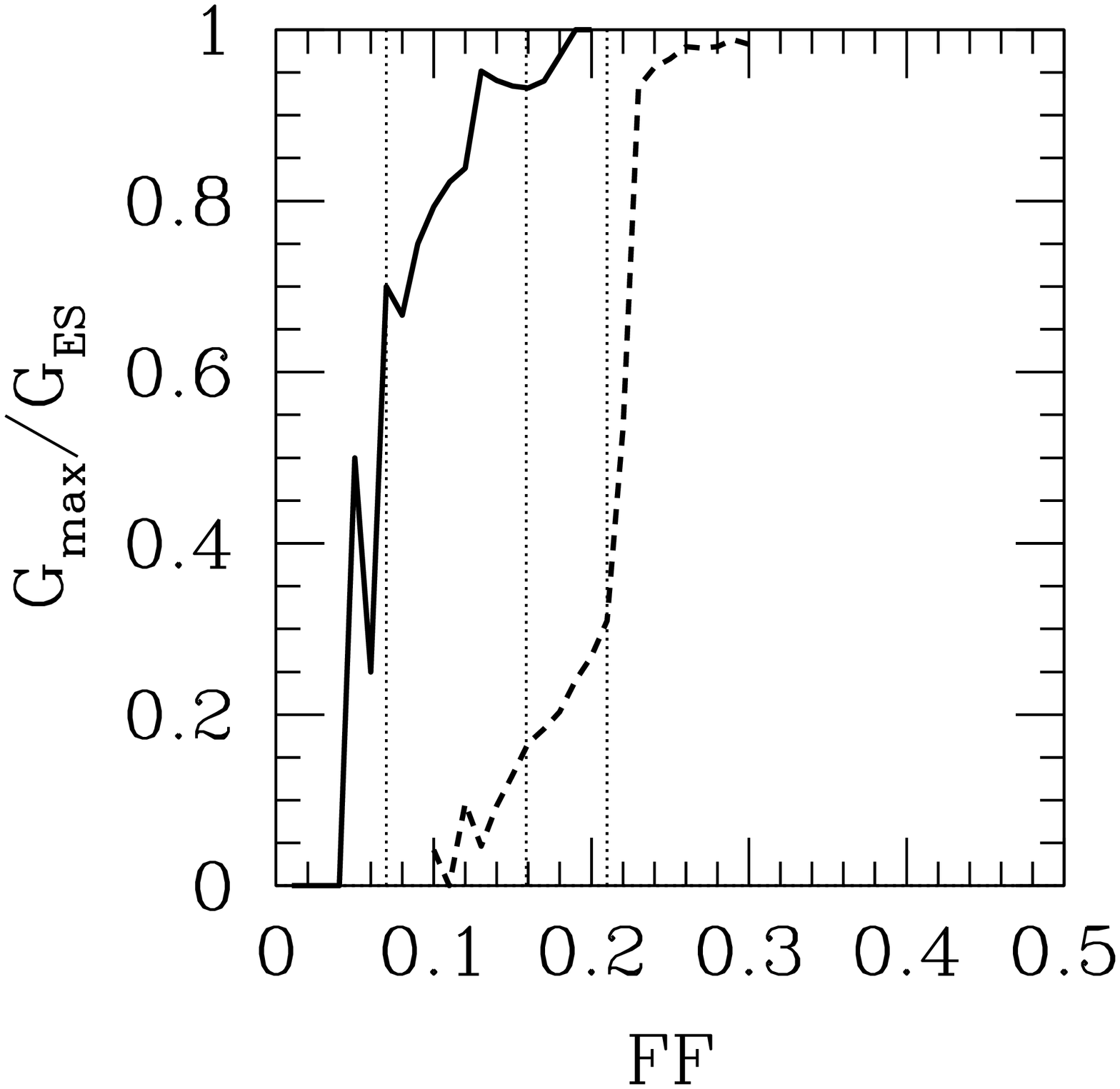}  
\end{minipage}\hfill
\caption{{\it Left panel}: the global genus is shown as a function of
  the filling factor for the density field smoothed with $L_s=5$
  $h^{-1}$ Mpc.  The half of the curve corresponding to high density
  thresholds is plotted as a function of $FF_C$ (solid line) while the
  other half corresponding to low density thresholds is plotted as a
  function of $FF_V$ (dashed line).  For comparison, the dotted line
  shows the Gaussian genus curve having the same amplitude.  The
  vertical dotted lines mark two percolation thresholds in the
  $\Lambda$CDM ($FF=FF_C \approx 0.07$ and $FF=FF_V \approx 0.22$) 
  and Gaussian
  field ($FF=FF_C=FF_V \approx 0.16$).  {\it Right panel}: the percolation
  transitions in the same density field as indicated; the genus of the
  largest supercluster (solid line) and largest void (dashed line).
  The vertical dotted lines mark the percolation thresholds similar to
  left panel.  }
\label{fig:gen_perc}
\end{figure}

A few years later \cite{gott86} (see also \cite{mel90} for a review)
suggested the genus statistic as a discriminator between various
models of large-scale structure. Although both percolation and genus
statistics characterize density fields and are sensitive to the
connectedness of the large scale structure, each carries 
significantly different information. 
It is important to remember that the genus refers to 
a surface which is the {\em interface} between over-dense and
under-dense regions (defined at a given density threshold).  An
interpretation of genus as a characteristic of a three dimensional
object (supercluster or void) bounded by this surface is not unique
and certainly non-trivial if the region has a complicated shape.  For
instance, both a full sphere and a doughnut with a bubble in its body have
genus of 0.
 
Figure \ref{fig:gen_perc} shows the genus curve in a slightly unusual form
(see also \cite{sss97}).  The half of the curve corresponding to high
density thresholds is shown as a function of the over-density filling factor
$FF_C$ (solid line) while the other half corresponding to low
thresholds is plotted as a function of the under-density filling
factor $FF_V$ (dashed line).  This allows to better illustrate the
deformations of the curve due to nonlinear effects. The Gaussian genus
curve is symmetric for positive and negative thresholds thus both 
parts of it overlap in Fig. \ref{fig:gen_perc} (dotted line).  The
vertical dashed lines mark three thresholds: the supercluster
percolation threshold at $FF=FF_C \approx 0.07$, the void percolation
threshold at $FF=FF_V \approx 0.22$ and the both percolation thresholds in
a Gaussian field at $FF=FF_C=FF_V\approx 0.16$.  

A marked decrease in the amplitude of the genus curve compared to the
Gaussian curve at small $FF$ is 
noticeable for both over-dense and under-dense excursion sets. 
(Small $FF \equiv$ high density for superclusters and low density
for voids.)
The global genus curve has no significant features at either percolation
threshold $FF_C \approx 0.07$ or $FF_V \approx 0.22$.  
The right panel of Fig. \ref{fig:gen_perc} shows the ratios
of the genus of the largest object $G_{max}$ to the global genus of the
excursion set  $G_{ES}$ for both superclusters (solid line) and voids
(dashed line). Both percolation
curves shown in the right panel would overlap in the case of a Gaussian
field (not shown) and demonstrate the percolation transition 
at $FF \approx 0.16$. The other indicators of the percolation transitions
(Eq. \ref{eq:lcs_def}) are in excellent agreement with 
the right panel of Fig. \ref{fig:gen_perc} which can be seen by
comparing  Fig. \ref{fig:gen_perc} with Figs. \ref{fig:perc_estim} 
and \ref{fig:perc_mf}.
The splitting of the percolation curves shown in Fig. \ref{fig:gen_perc}
(right panel) as well as in Fig. \ref{fig:perc_mf}
clearly demonstrates non-Gaussianity of the density field.
\begin{figure}
\centering
\centerline{
  \includegraphics[width=3.3in]{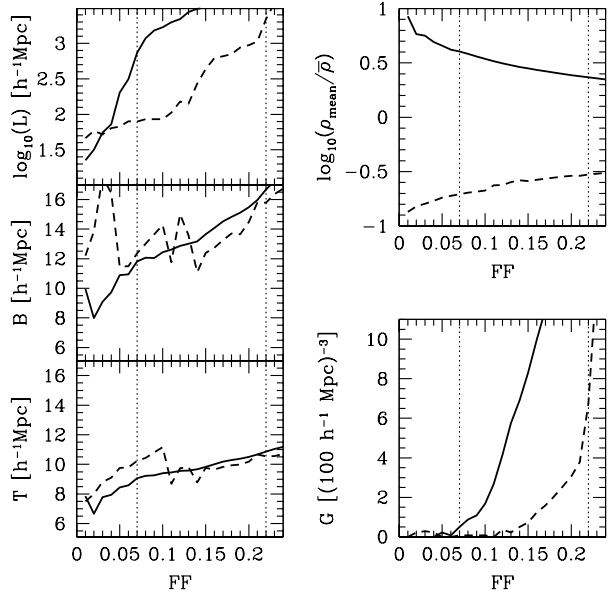} }
\caption{The length, breadth, thickness, mean density, and  
  genus of the largest (by volume) supercluster (solid lines) and void
  (dashed line) in the \L CDM model smoothed with $L_s=5~h^{-1}$ Mpc
  are shown as a function of of $FF_C$ and $FF_V$ respectively.
  Vertical dotted lines show the supercluster ($FF= FF_C \approx
  0.07$) and void ($FF=FF_V \approx 0.22$) percolation thresholds.  }
\label{fig:perc_lbtg}
\end{figure}

\begin{figure}
\begin{minipage}[t]{.99\linewidth}
\centering
  \includegraphics[width=4cm]{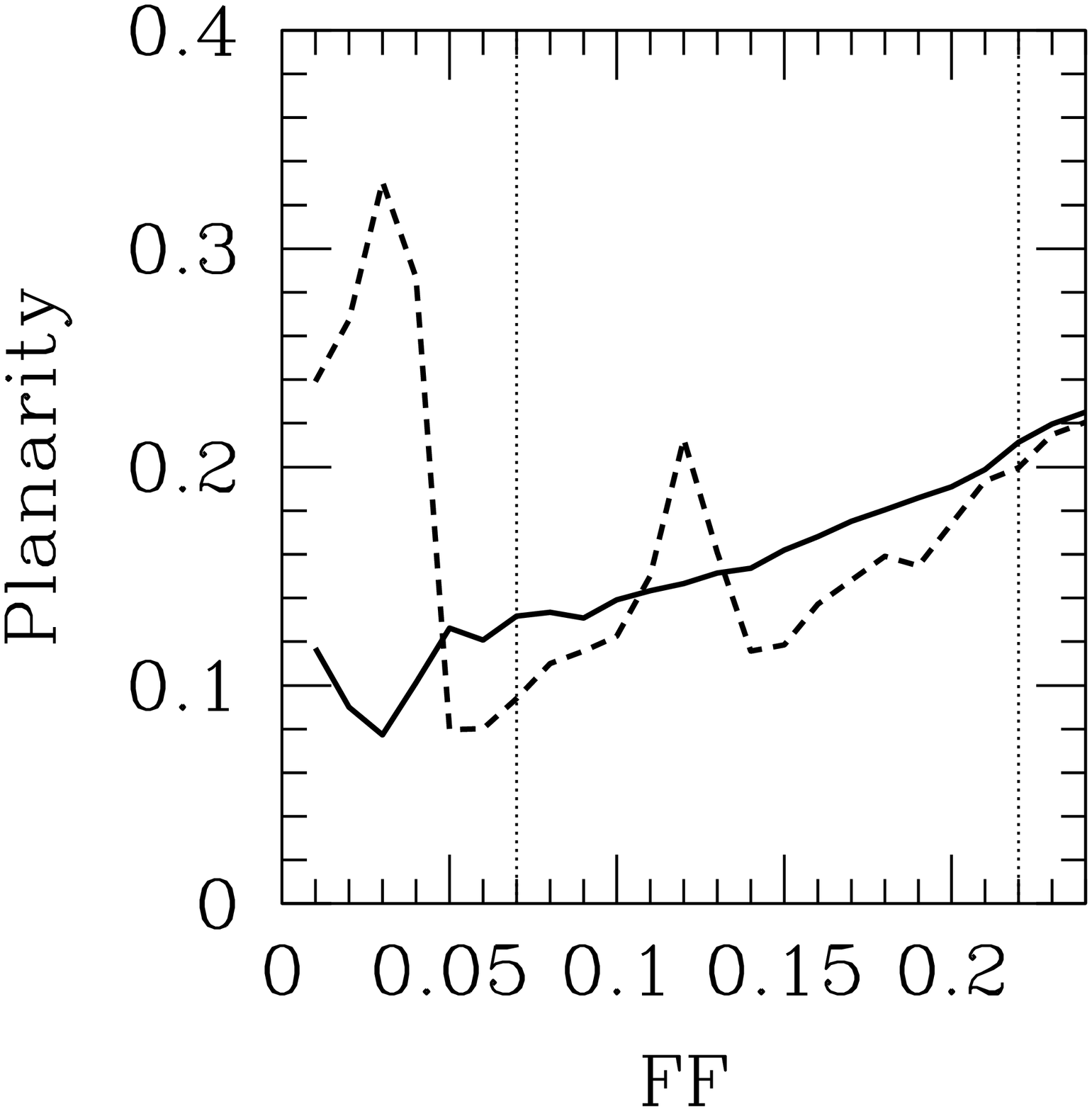}
  \includegraphics[width=4cm]{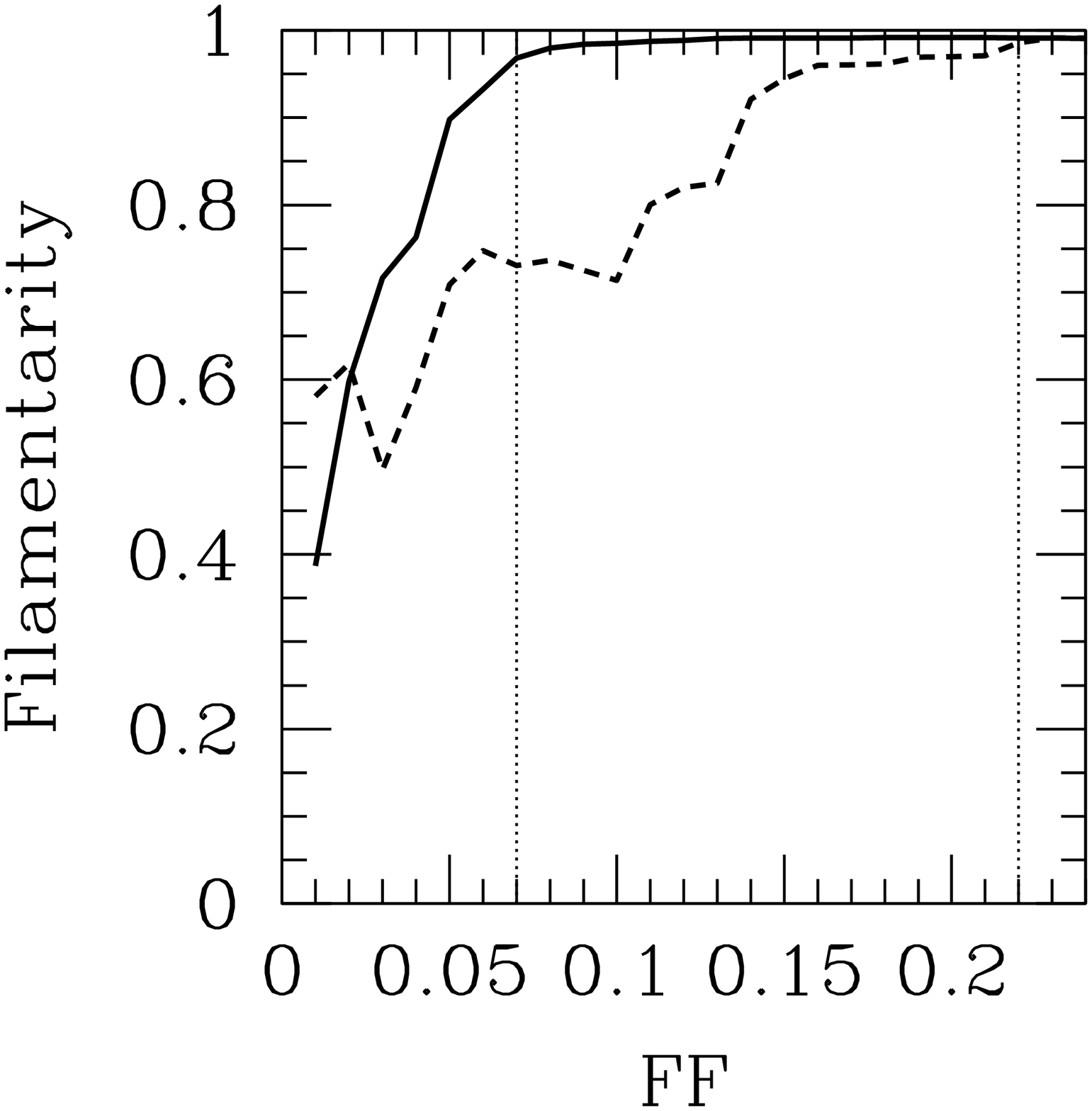}  
\end{minipage}\hfill
\caption{Planarity and filamentarity of the largest structures as a function of
  the filling factor. The largest supercluster and void are shown by
  solid and dashed lines respectively.  Vertical dotted lines show the
  supercluster ($FF=FF_C \approx 0.07$) and void ($FF=FF_V \approx 0.22$)
  percolation thresholds.}
\label{fig:perc_pf}
\end{figure}

After percolation takes place the genus of the largest object
considerably increases (Fig. \ref{fig:perc_lbtg}) which manifests its
complex shape.  The length and therefore filamentarity of the largest
object radically changes after percolation takes place (see Fig.
\ref{fig:perc_lbtg} and \ref{fig:perc_pf}) and cannot be easily
interpreted.  On the other hand, the breadth and thickness, (and consequently
also the planarity), grow much more 
gradually with the growth of the corresponding filling
factor and in this sense are
similar to non-percolating objects. 
The {\em mean density} of
the percolating supercluster at the percolation transition is
$\bar{\rho}_{\rm SC}\approx 4 {\bar\rho}$ so that 
$\delta_{\rm SC}\approx 3$. The corresponding value for the percolating
void is 
$\bar{\rho}_{\rm V} \approx 0.3$ and $\delta_{\rm V}\approx - 0.7$.
Both $\delta_{\rm SC}$ and $\delta_{\rm V}$ are 
significantly different
from the percolation threshold for superclusters $(\delta_{\rm C}
= 1.8 {\bar\rho})$ and voids ($\delta_{\rm V} = -0.5$) respectively
(see Figure \ref{fig:perc_lbtg}).
\section{Individual Superclusters and Voids}
\begin{figure}
\begin{minipage}[t]{.99\linewidth}
\centering
  \includegraphics[width=4cm]{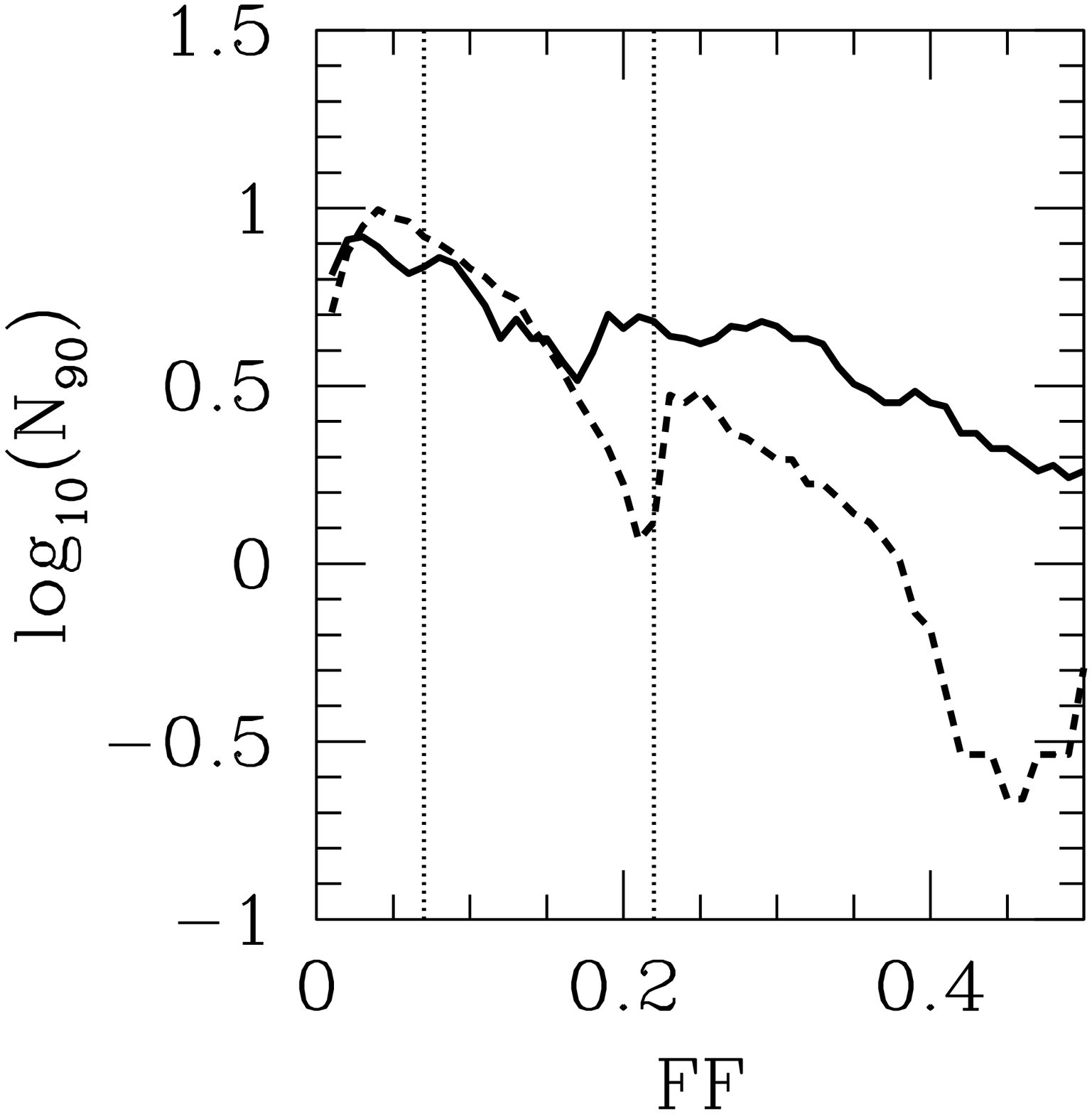}
  \includegraphics[width=4cm]{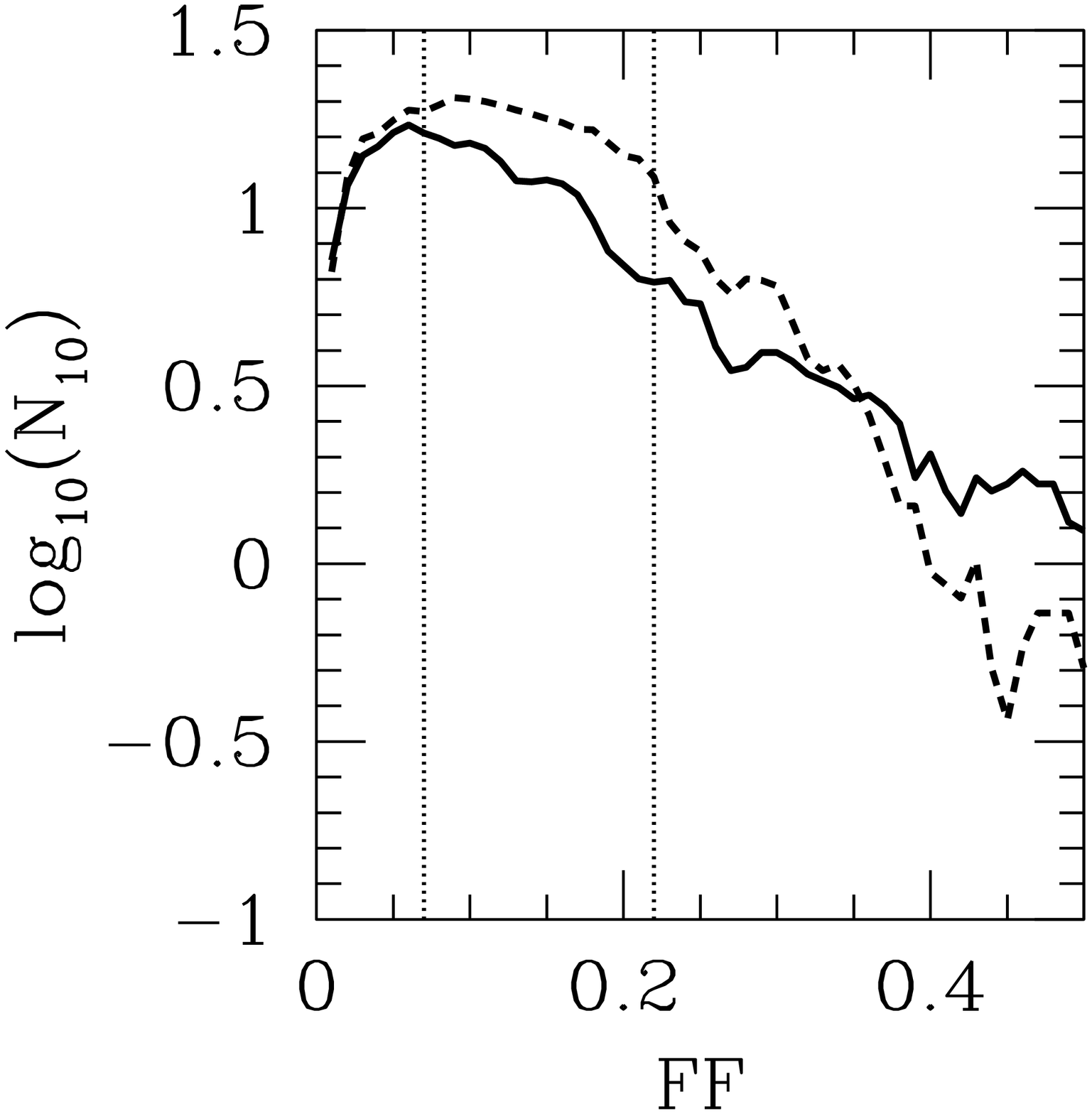}  
\end{minipage}\hfill
\caption{{\it Left panel}: the number density 
 (in (100 h$^{-1}$ Mpc)$^{-3}$ units)
  of the most massive superclusters making 90\% of all {\it mass}
  contained in non-percolating superclusters as a function of the filling
  factor $FF_C$ is shown by solid line. The number density of the
  largest voids making 90\% of all {\it volume} contained in non-percolating
  voids as a function of the filling factor $FF_V$ is shown by dashed
  line.  {\it Right panel}: the number density of the least massive
  superclusters making 10\% of all {\it mass} contained in non-percolating
  superclusters as a function of the filling factor $FF_C$ is shown by
  solid line. The number density of the smallest voids making 10\% of
  all {\it volume} contained in non-percolating voids as a function of the
  filling factor $FF_V$ is shown by dashed line. Two vertical dotted
  lines mark percolation thresholds for superclusters ($FF_C \approx
  0.07$) and voids($FF_V \approx 0.22$) }
\label{fig:number}
\end{figure}
In this section we carry out comparative statistical analysis of
physical and geometrical parameters of superclusters and voids. We
stress that the largest supercluster and the largest void both are
excluded from this analysis.  Before percolation transition they both 
are extreme outliers and after percolation they have nothing
in common with the other objects.  The total number of individual objects
is shown in Fig. \ref{fig:number}.  There are numerous small objects
among both superclusters and voids which dominate by numbers at every
threshold (right panel of Fig. \ref{fig:number}).  
Including them into the statistical analysis along with large
objects would seriously affect all the parameters.  One way to deal
with this problem would be the computation of weighted parameters,
\ie by mass for superclusters and by volume for voids.  This actually
corresponds to computing one or a few moments of the probability
distribution function which also may be misleading because the
distribution functions are strongly non-Gaussian.  We analyze only the
the most massive superclusters making 90\% of all mass in
non-percolating over-dense objects and most voluminous voids 
making 90\% of total volume in non-percolating under-dense objects.  
The smallest over-dense and
under-dense objects are excluded from the analysis.  Figure
\ref{fig:number} shows the number density in (100 h$^{-1}$ Mpc)$^{-3}$
units for both large (left panel) and small (right panel) objects.
Despite the smallest objects make only 10\% of mass or volume
they are more numerous then the largest objects making the most of
mass or volume.
\subsection{Masses, volumes and mean densities}
\begin{figure}
\centering
\centerline{
  \includegraphics[width=3.3in]{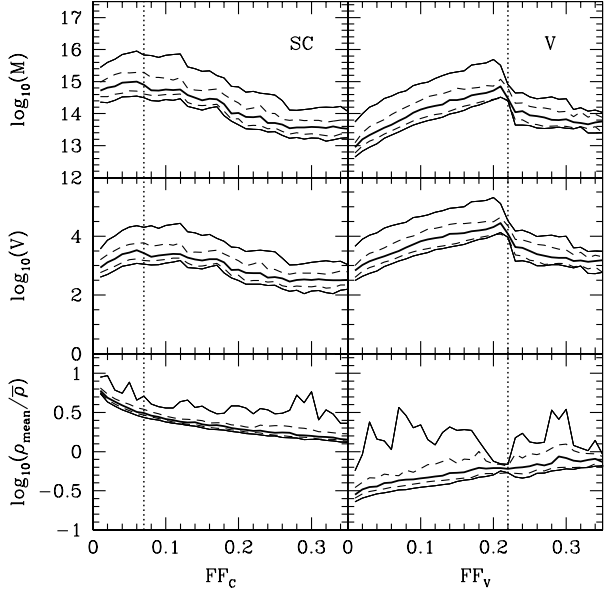} }
\caption{Mass (in solar masses), volume (in (h$^{-1}$ Mpc)$^3$), 
  and mean density (\ie $\rho_{mean} =(M/V)/\bar{\rho}$) of all but
  the largest structure are shown as a function of the volume filling
  factor $FF_C$.  Thick solid line is the median of the distribution,
  dashed lines mark 25-75\% interval, and thin solid lines show the
  third largest and third smallest value in the distribution. Vertical
  dotted lines show the percolation thresholds in over-dense and
  under-dense excursion sets. The supercluster and void  parameters are shown 
  in the right and left panels respectively.}
\label{fig:mvrho}
\end{figure}
We begin with the analysis of masses, volumes and mean densities
\b
\rho_{mean}\equiv \frac{M}{V}
\e
which are probably the most important factors among those that determine
the visual impression. They are also the least ambiguous.

Figure \ref{fig:mvrho} shows the masses, volumes and mean densities of
superclusters (left) and voids (right) at various thresholds
parameterized by the corresponding filling factor.  The thick solid
lines show the median of the distribution, the dashed lines show the
50\% interval (25\% largest and 25\% smallest are beyond this
interval), and thin solid lines show the 95\% interval.  The intervals
are highly asymmetric with respect to the median which indicates 
the non-Gaussianity of all distribution functions.  
Every distribution function  has a long tail at large values. 
Both superclusters and voids have largest masses and volumes
around the corresponding percolation threshold although the maximum 
is more distinct in the case of voids. 
As one might anticipate, voids are less massive and
more voluminous than superclusters. However, the difference is not
huge at the chosen smoothing scale.  The median mass reaches a maximum 
about $10^{15}\ M_{\odot}$ for superclusters and $10^{14.8}\ M_{\odot}$ for 
voids.  The median volume in the corresponding maximum is about $3\times
10^3\ (h^{-1} Mpc)^3$ for superclusters and about $3\times 10^4\ 
(h^{-1} Mpc)^3$ for voids.  The mean densities of the superclusters
are well above the mean density before percolation and gradually
decrease after percolation takes place however remaining above the mean
density in the universe.  
As expected the mean density of voids increases with the
growth of $FF_V$ corresponding to the growth of the density threshold.
It is somewhat surprising that the volumes of superclusters do not
change much with the threshold, the masses change more but not 
greater than by an order of magnitude.
\subsection{Sizes and Shapes}
\begin{figure}
  \centering \centerline{\includegraphics[width=3.3in]{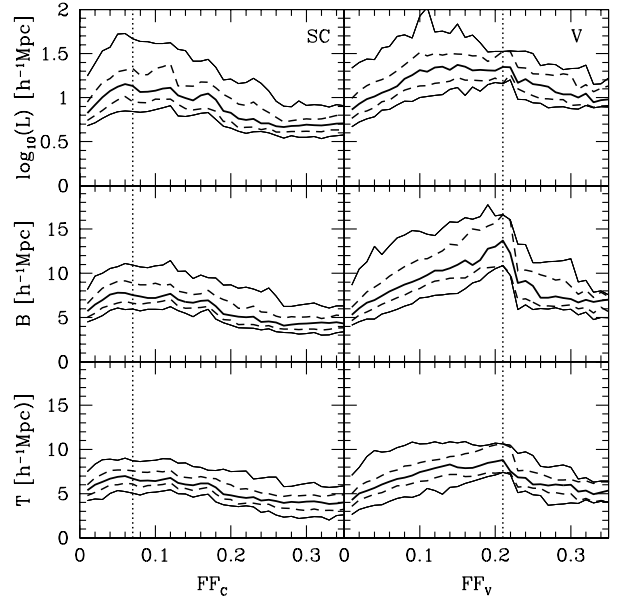}}
\caption{The length $L$, breadth $B$,  and thickness $T$ of all but the 
  largest object is shown as a function of the 
  corresponding filling factor.
  The notations are as in Fig. \ref{fig:mvrho}. }
\label{fig:lbt}
\end{figure}
Three characteristic sizes and shapes of superclusters and voids can
be estimated from \mf ~ of every object (eq. \ref{eq:TBL} and
 \ref{eq:shapefinder}).  Figure \ref{fig:lbt} shows the median and 50\%  
and 95\% intervals for the length, breadth and thickness of the
superclusters and voids. The sizes of superclusters are shown as 
a function of $FF_C$, while the sizes of voids as a function of $FF_V$. 
It is surprising that the median thickness of
superclusters depends on the threshold so weakly; it is within 4$-$6
h$^{-1}$Mpc interval for a range of thresholds between $0 \lleq \delta
\lleq 6$.  This may indicate that the actual thickness of
superclusters is significantly smaller and the measured values reflect
the width of the smoothing window. The breadth of superclusters is not
much larger than the thickness and it is likely that this quantity is also
affected by the
width of the filtering window.  Voids are a little fatter than superclusters
and their
median thickness reaches about 9 h$^{-1}$ Mpc at the percolation
threshold.  Interestingly voids are also 
wider and longer
than superclusters (please note the logarithmic scale used for length).
Recalling that the size parameters are normalized to the radius of the
sphere rather than to diameter we conclude that the longest 25\% of
superclusters are longer than about 50 h$^{-1}$ Mpc and 25\% of voids are
longer than about 60 h$^{-1}$ Mpc.
\begin{figure}
\centering
\centerline{
  \includegraphics[width=3.3in]{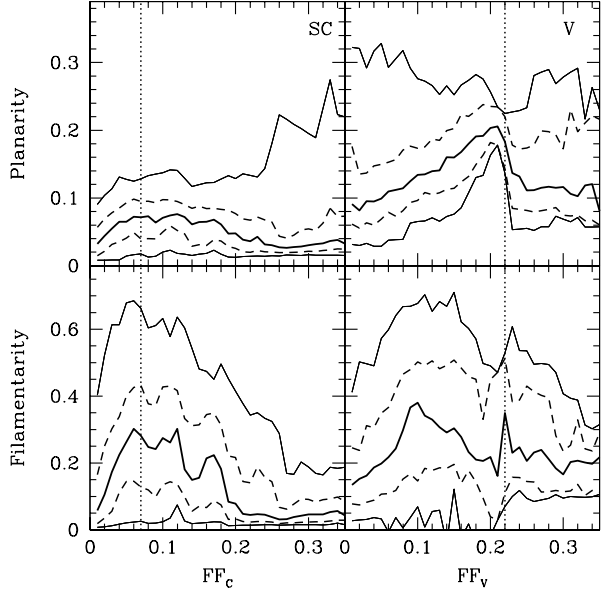} }
\caption{The planarity, P and filamentarity, F of all but the 
  largest object is shown as a function of the 
corresponding filling factor.
  The notations are as in Fig. \ref{fig:mvrho}.}
\label{fig:pf}
\end{figure}

The shape parameters of superclusters and voids 
are shown in Fig. \ref{fig:pf}.  The median
planarities of superclusters are small which means that in the \L CDM
universe the dark matter density field smoothed with 5h$^{-1}$Mpc
typically has no pancake-like superclusters with the diameters greater
than about 10 h$^{-1}$ Mpc. The outliers can reach planarity $P\approx
0.3$ corresponding to the ratio $B/T \approx 1.4$ which is not 
large either.  In addition, it happens only at quite large filling factors
where the density threshold is quite low and only very few
superclusters are left (see Fig. \ref{fig:number}).  
Voids show an
opposite trend to superclusters: voids are significantly more planar than
superclusters and the largest planarities in voids occur at small
filling factors.  Filamentarities are significantly higher for both
over-dense and under-dense objects: median values peak at about 0.35
($L/B \approx 2 $). The outliers could be considerably more elongated
$F > 0.7$ (\ie $L/B > 6 $). As we shall see in the next section, the
more massive a supercluster or more voluminous a void, the greater is
its tendency to be filamentary.
\subsection{Correlations between morphological and physical parameters}
\begin{figure}
  \centering \centerline{ \includegraphics[width=3.3in]{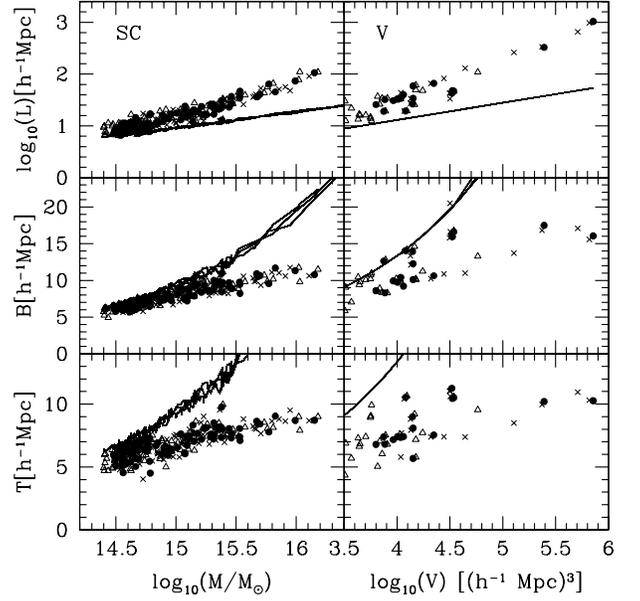} }
\caption{The length, breadth, and thickness versus mass for superclusters
  and versus volume for voids at percolation. Solid circles show the
  relation at percolation thresholds: $FF_C=0.07$ for superclusters
  and $FF_V=0.22$ for voids.  Crosses show the parameters before
  percolation ($FF_C=0.06$ for superclusters and $FF_V=0.21$ for
  voids) and empty triangles after percolation ($FF_C=0.08$ for
  superclusters and $FF_V=0.23$ for voids).  Solid lines show the
  radius of the sphere having the same volume as the corresponding object. 
  Note the logarithmic scale used for the length. Three lines correspond to
three different thresholds.}
\label{fig:lbt_dots}
\end{figure}
Both superclusters and voids reach their largest sizes near their respective
percolation thresholds: $FF_C \approx 0.07$ for superclusters and 
$FF_V \approx 0.22$ for voids. Figure \ref{fig:lbt_dots}
shows the scatter plots of three characteristic sizes ($L$, $B$, and
$T$) versus mass for superclusters and versus volume for voids.  The
combined plots are made for $FF_C=0.06$, $0.07$ and $0.08$ for
superclusters and for $FF_V=0.21$, $0.22$ and $0.23$ for voids.  The
solid lines show the radius of a sphere having the same
volume as a given object ($R=(3 V/4 \pi)^{1/3}$). 
All three sizes show a significant
correlation with the mass: the greater the mass the greater the thickness,
breadth and length. The thickness and breadth approximately double 
their values and length grows
over an order of magnitude when mass increases from about
$10^{14.5}\ M_{\odot}$ to $10^{16.5}\ M_{\odot}$.  Both the thickness
and breadth are considerably smaller than the radius $R$ of a sphere
having similar volume for large superclusters 
($M \ggeq 10^{15}\ M_{\odot}$) as well as for large voids 
($V \ggeq 10^{4}\ (h^{-1}\ Mpc)^3$).
On the other hand the length is considerably greater than $R$. This is
another manifestation of anisotropies of the large-scale objects.

\begin{figure}
\centering
\centerline{
  \includegraphics[width=3.3in]{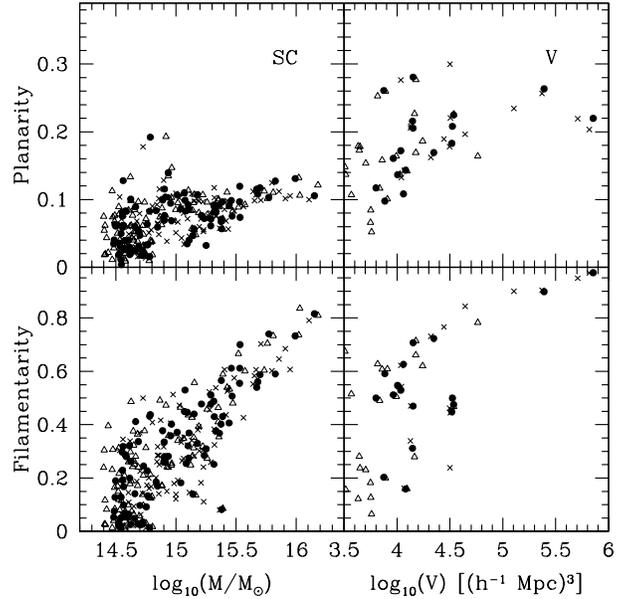} }
\caption{The planarity and filamentarity vs mass (for superclusters)
  and vs volume (for voids) at percolation.  Notations are as in Fig
  \ref{fig:lbt_dots}.}
\label{fig:pf_dots}
\end{figure}

\begin{figure}
\begin{minipage}[t]{.99\linewidth}
\centering
  \includegraphics[width=4cm]{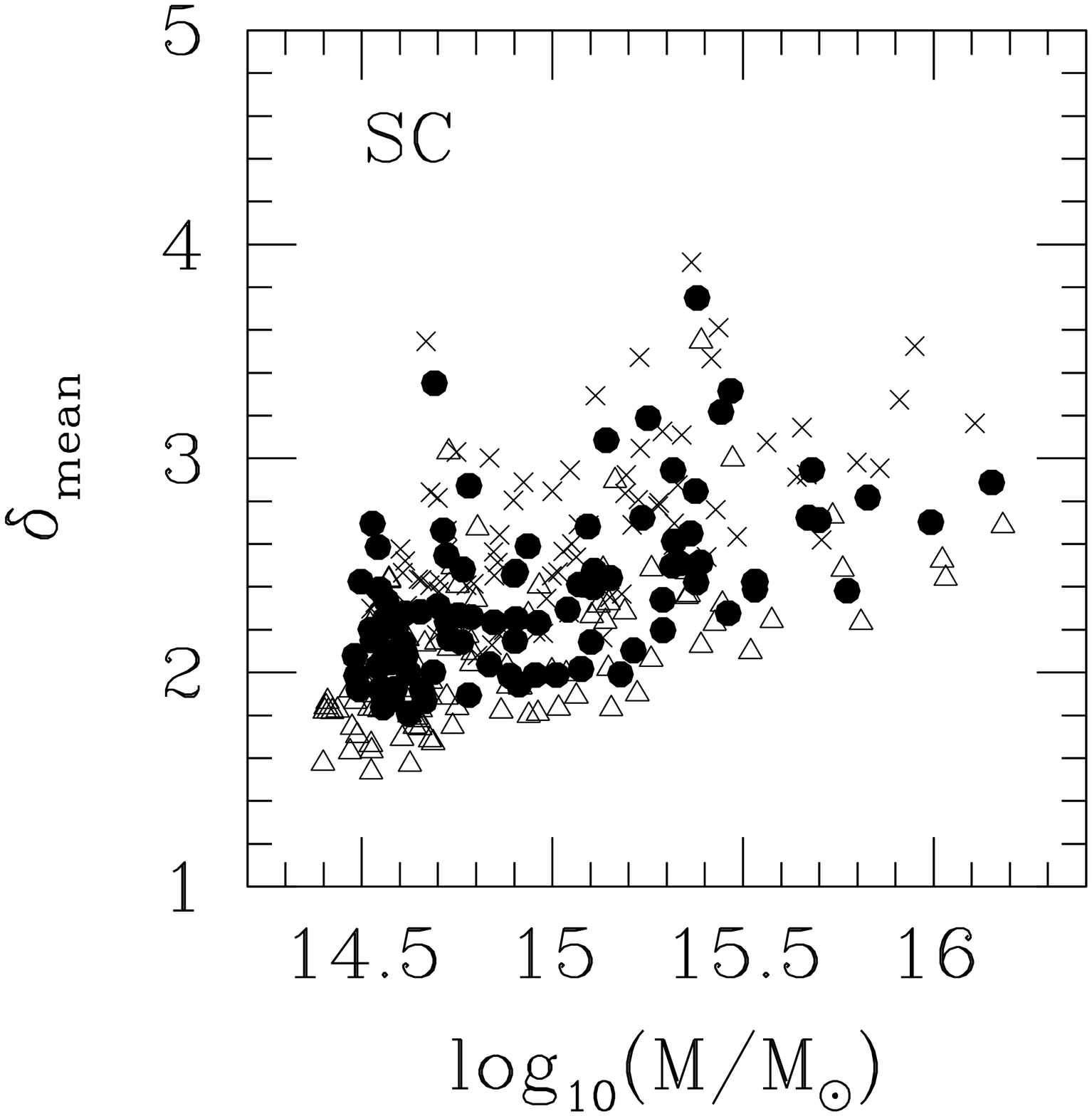}
  \includegraphics[width=4cm]{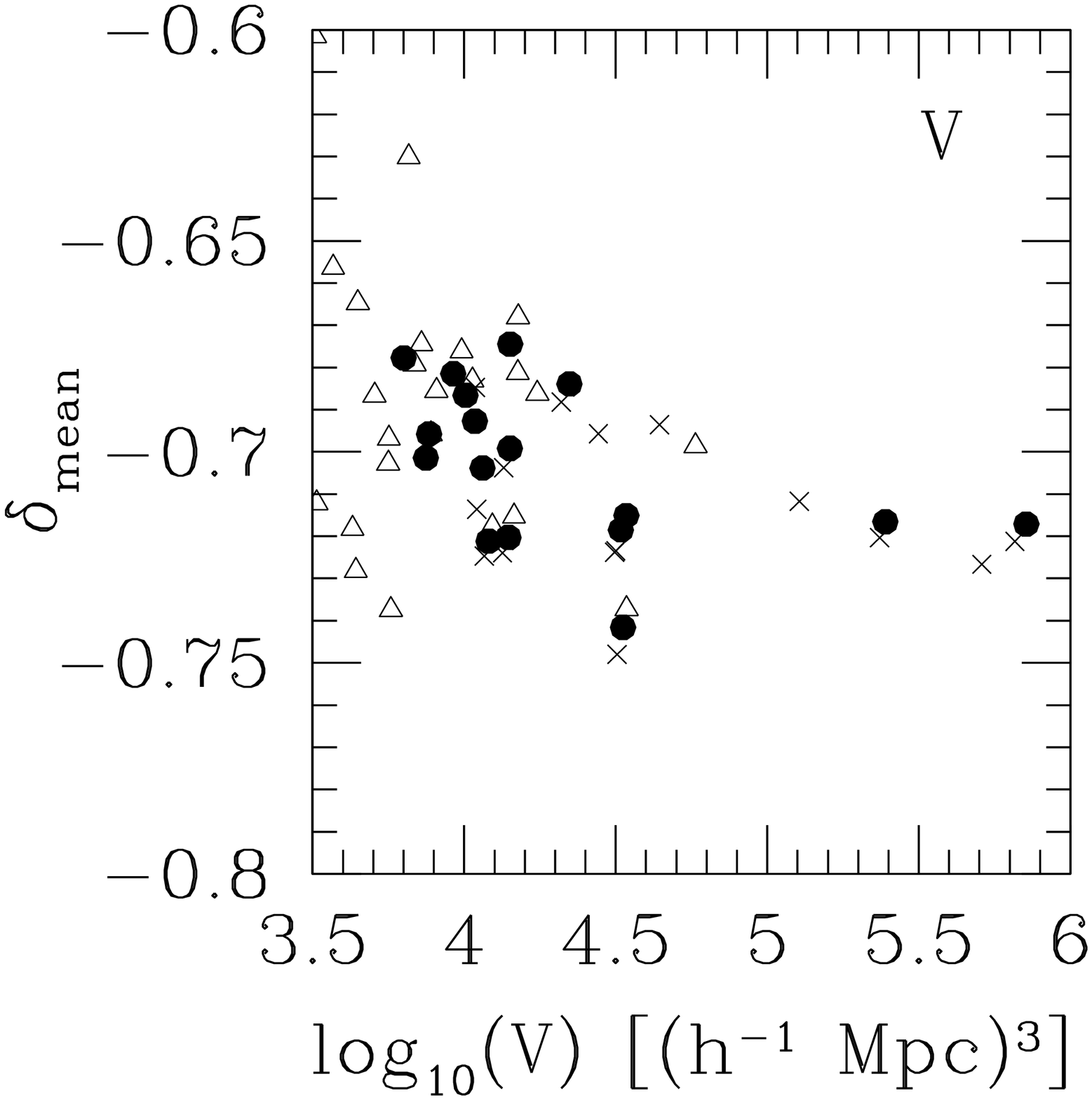}
\end{minipage}\hfill
\caption{{\it Right panel}: correlation of the mean density contrast 
  ($\delta_{mean}=M/(\bar{\rho}V) -1$) of a supercluster with its mass
  at percolation. {\it Left panel}: correlation of the mean density contrast
  of a void with its volume at percolation.  Notations are as in Fig
  \ref{fig:lbt_dots}.}
\label{fig:del_mv}
\end{figure}

The corresponding plots of planarities and filamentarities show a
similar correlation: the larger the mass of a supercluster or the larger
the volume of a void, the greater its planarity and filamentarity (Fig.
\ref{fig:pf_dots}). One can clearly see that the largest objects
(superclusters with $M > 10^{15}\ M_{\odot}$ and voids with $V >
10^{4}\ (h^{-1}Mpc)^3$ are the most anisotropic large-scale objects.
Note that voids display a higher level of planarity when
compared to superclusters.
Indeed, one of the most noticeable results of this analysis is 
evidence that voids defined near the onset of percolation of the
under-dense excursion set are {\em significantly} non-spherical.
Finally, the larger the mass of a supercluster the greater its mean
density and the more voluminous a void the lower its mean density
(Fig. \ref{fig:del_mv}).
\section{SUBSTRUCTURE IN SUPERCLUSTERS AND VOIDS}
\begin{figure}
\begin{minipage}[t]{.99\linewidth}
\centering
  \includegraphics[width=4cm]{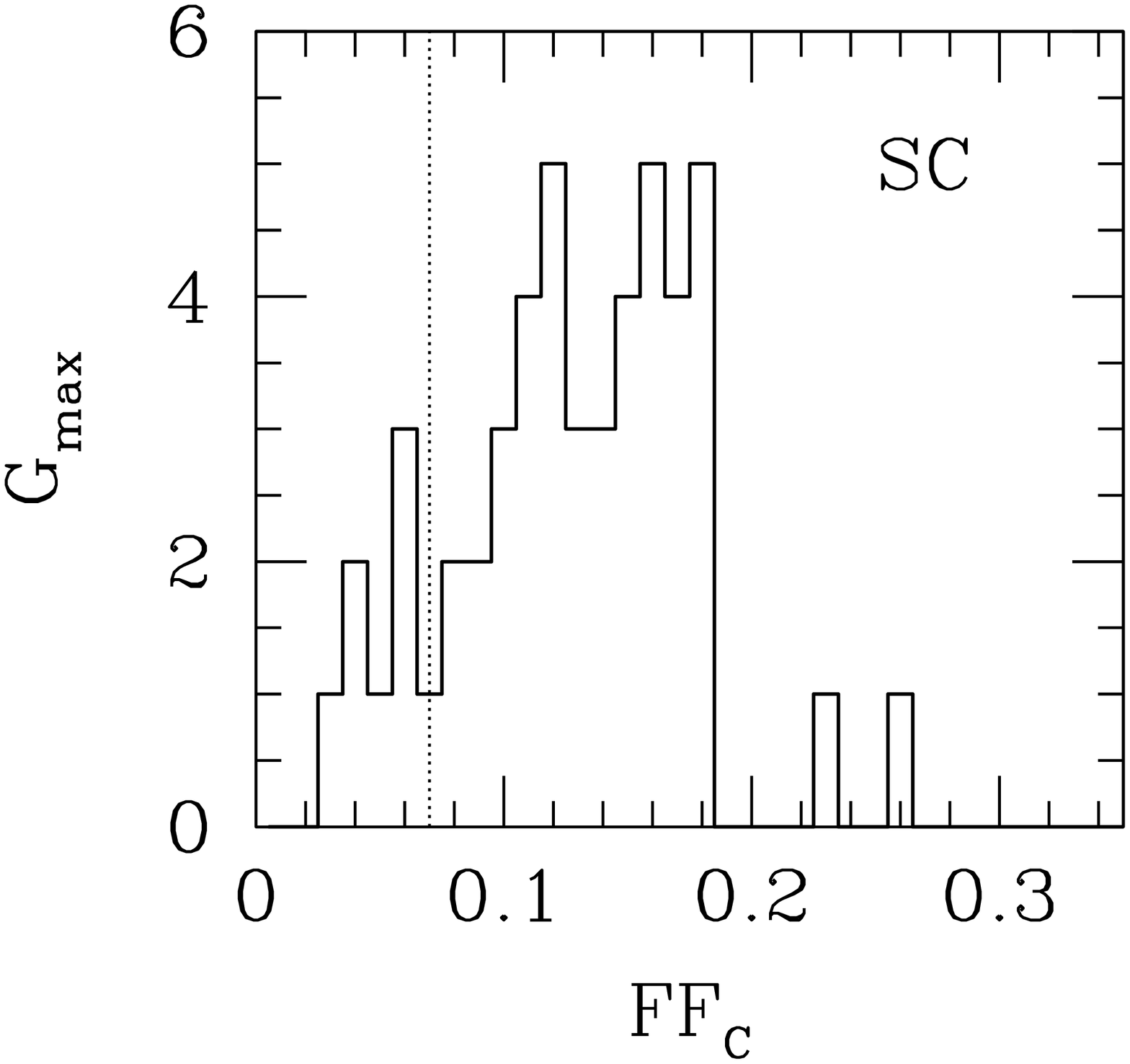}
  \includegraphics[width=4cm]{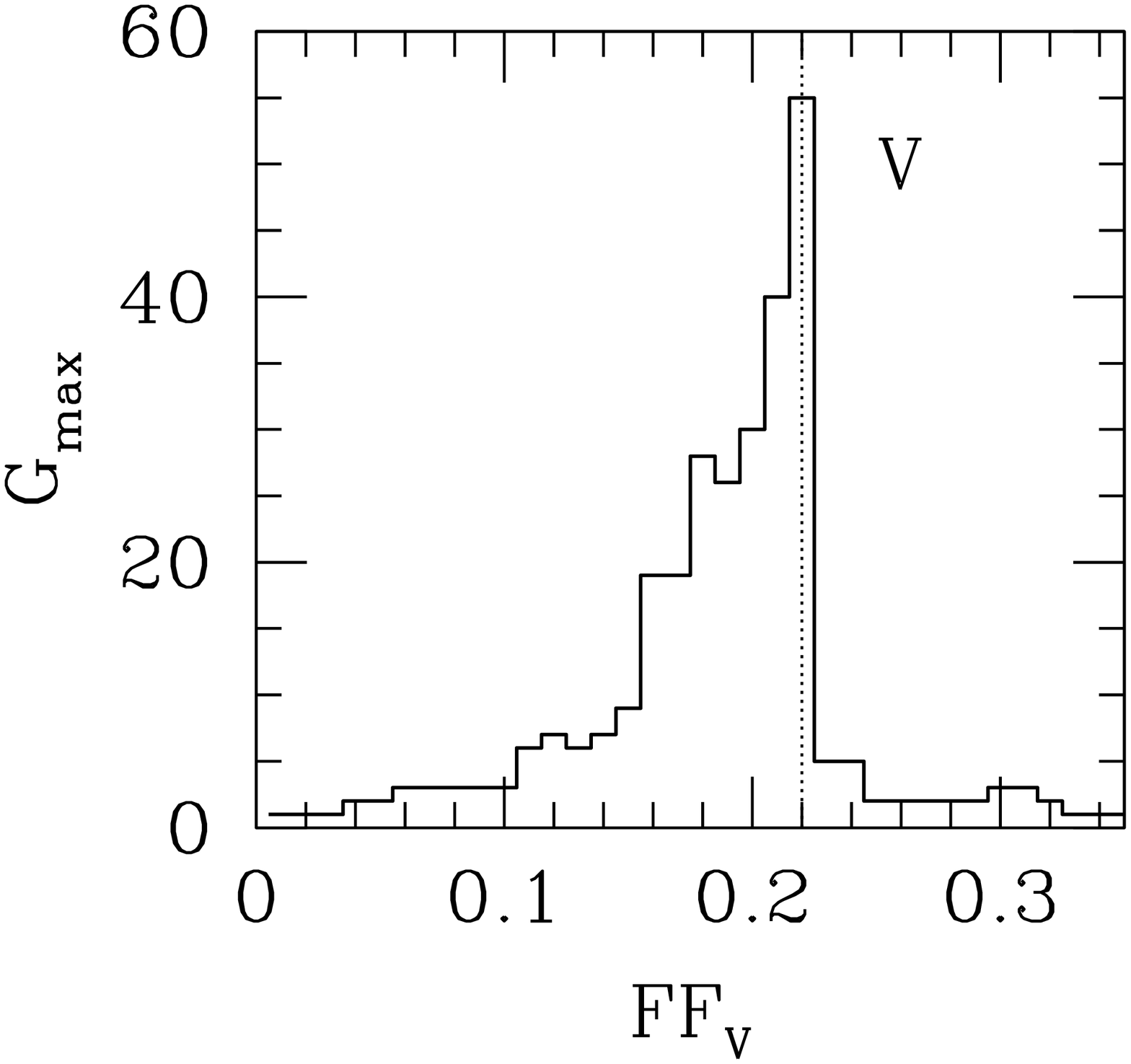}  
\end{minipage}\hfill
\caption{The maximum of genus as a function of the filling factor.
  Left panel: the maximum genus in isolated superclusters.  Right
  panel: the maximum genus in isolated voids.  Note the difference of
  the scales on the vertical axis.}
\label{fig:gen_max}
\end{figure}
Visual inspection shows that superclusters and voids have noticeable
substructure. Isolated voids inside superclusters and isolated
clusters and superclusters inside voids are obvious examples.  Theory
\citep{kof_etal92,sss94} and targeted N-body simulations
\citep{bea_etal91,gottl_etal03} show that voids are also filled with
smaller filaments of high density. In the \L CDM universe these
filaments are strong enough to survive smoothing on a scale $L_s=5\ 
h^{-1}Mpc$.  Figure \ref{fig:gen_max} shows the maximum of the genus
of superclusters and voids. In order to interpret it properly we need
to explain how to interpret the genus of an object.  A standard
interpretation of genus in cosmological literature (see \eg \citet{mel90}) 
says
\b
\label{eq:genus}
G=\mbox{(number of holes)-(number of isolated regions)}+1, 
\e
this requires additional clarifications.  First, ``holes'' stand for
fairly complex mathematical objects that do not always coincide with
the simple minded meaning of the word.  For instance, a cavity in a
tooth is not a hole, a bubble in a glass of champagne is not a hole
either, but a tunnel through a mountain is a hole. Second, ``number of
isolated regions'' means the number of isolated pieces of the surface
that define the boundaries of the object.  For instance, a region
bounded by two nested spheres (a thick spherical shell) has two
disconnected pieces of the boundary surface and no tunnels, therefore
its genus is $-$1.  A torus has surface in one piece but it has a
tunnel, thus its genus equals +1.  Consider an object topologically
homeomorphic to a sphere (G=0), \ie only one boundary surface and no
tunnels. Adding a tunnel increases the genus by one and adding a
bubble inside the body reduces the genus by one.
Thus, a doughnut with a bubble of air in its body has genus 0, exactly as
a full sphere.  

Figure \ref{fig:gen_max} shows that the genus of
an isolated supercluster can be as large as five. It means that there
are {\em at least} five tunnels through the supecluster.  The number
of tunnels could be even greater if it also had a few smaller 
isolated voids inside. The vast majority of
superclusters are topologically isomorphic to a sphere \ie, $G=0$,
but the most massive ones have genus greater than 
unity. A supercluster with genus of unity is homeomorphic
to a doughnut and one with genus of two to a pretzel.
The substructure of voids is considerably more complex.  The largest
genus of voids detected in this simulation is 55, therefore at least
55 filaments span through the void !

\begin{figure}
\begin{minipage}[t]{.99\linewidth}
\centering
  \includegraphics[width=4cm]{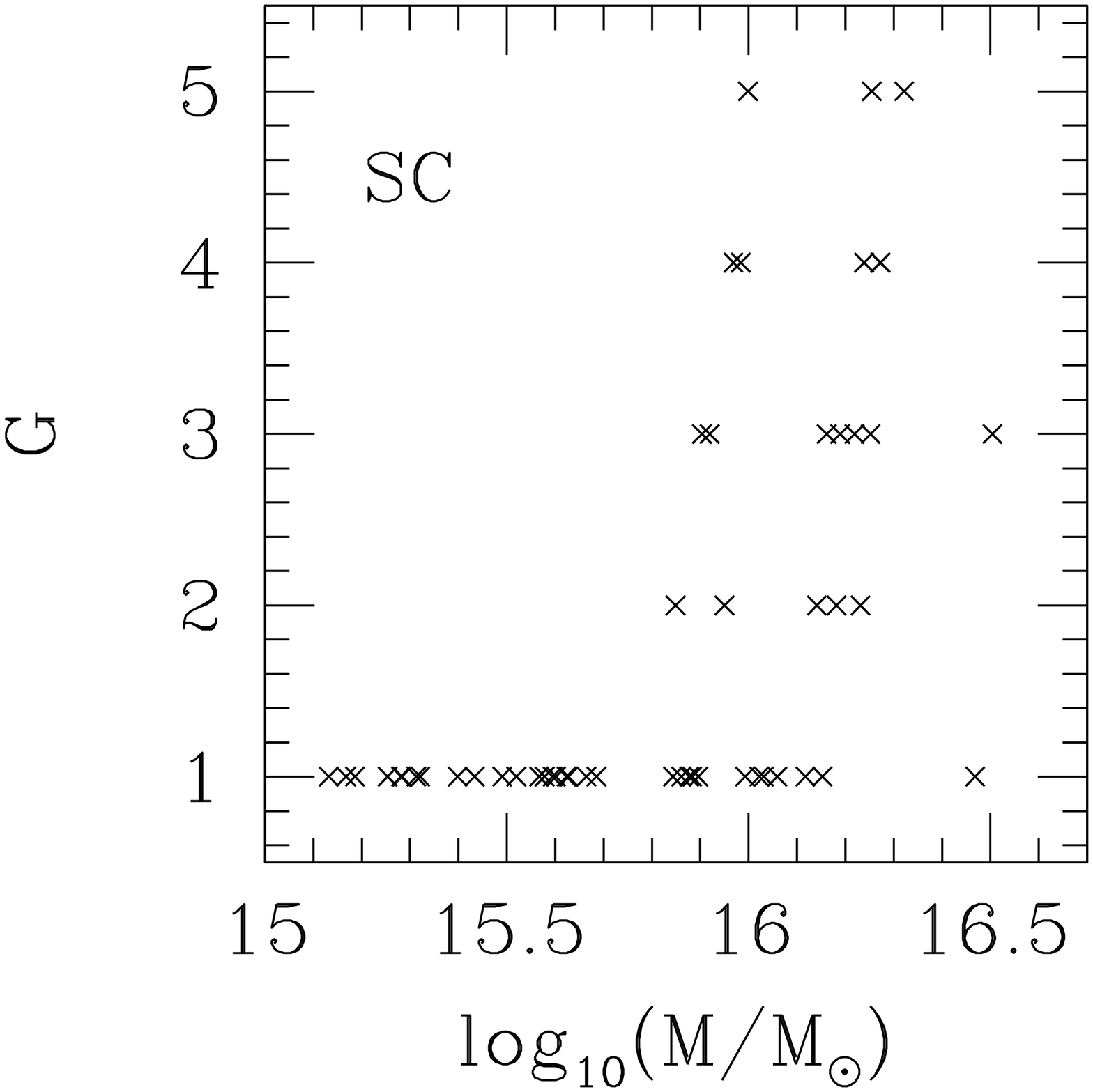}
  \includegraphics[width=4cm]{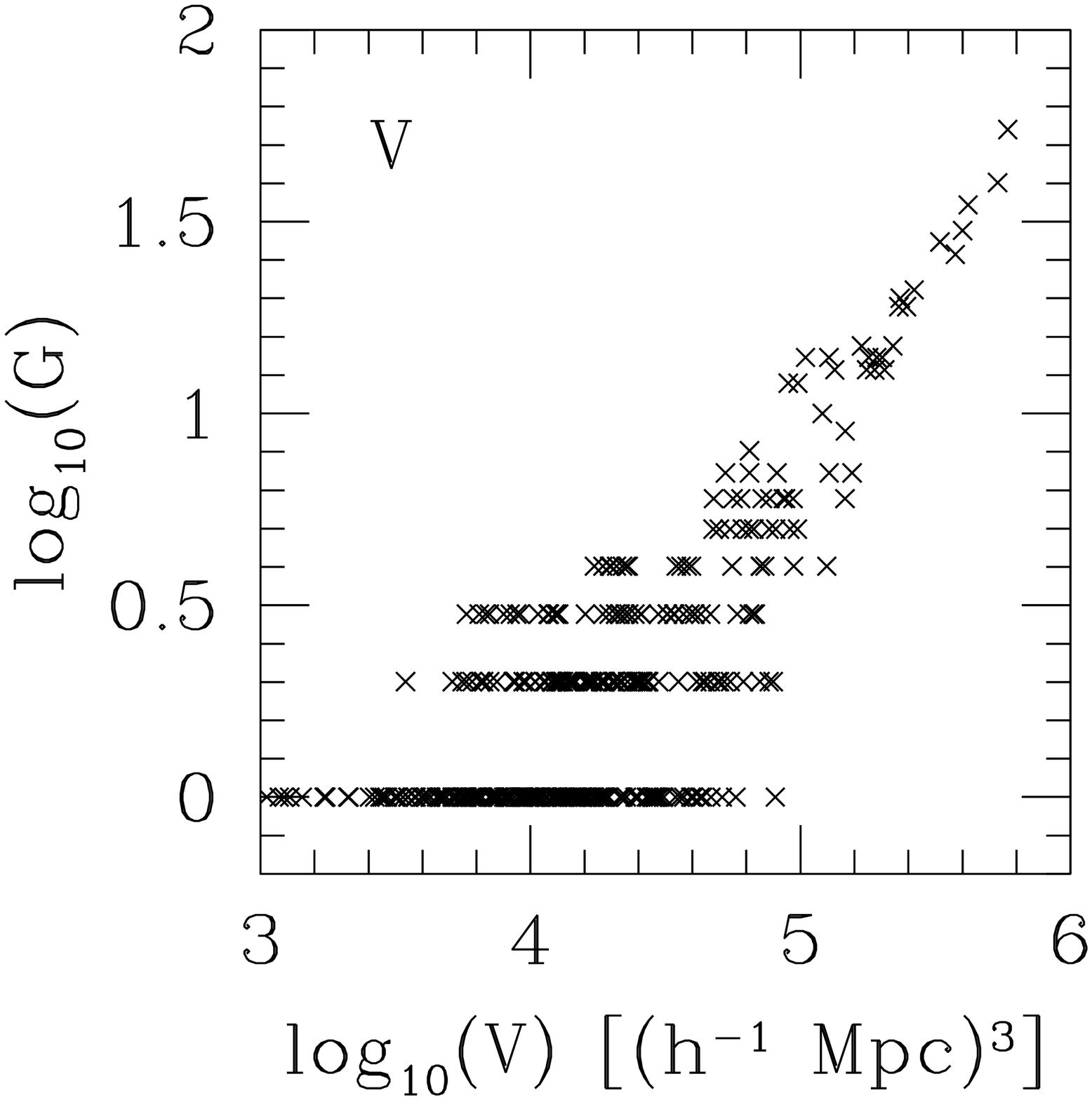}  
\end{minipage}\hfill
\caption{The genus-mass relation is shown for isolated superclusters
  in the left hand side panel. The genus-volume relation is shown for
  voids on the right.  Every isolated supercluster and void having
  genus greater than zero at all density thresholds is shown.  The
  mass is given in the solar units and the volume in (h$^{-1}$
  Mpc)$^3$.  Note the difference of the scales on the vertical axis.}
\label{fig:gen_mv}
\end{figure}

Figure \ref{fig:gen_mv} shows the correlation between genus and mass
(for superclusters) or volume (for voids). Largest superclusters
 $M \ggeq 10^{15}\ M_{\odot}$ have a nontrivial topology. Generally
the more massive supercluster the greater chance of complex topology.   
Voids display even stronger correlation between genus
and volume. 
\section{Symmetry of Global Minkowski Functionals}
The global \mf ~were introduced in cosmology by \cite{meckwag94} and
since then have been used in many studies, see \eg 
\citet{schmal99,sh-s-s-sh03}.  Since the \mf ~are additive\footnote{
Note, genus is not additive but the Euler characteristic is additive.}
we
obtain them by simple summation over all objects at every density
threshold. As discussed in \cite{sh-s-s-sh03}) computing the global
\mf ~poses some problem in the case of periodic boundary conditions.
However, as shown in Appendix it can be solved if \mf ~ are computed 
for every supercluster and void at a given threshold.

\begin{figure}
\centerline{
  \includegraphics[width=3.3in]{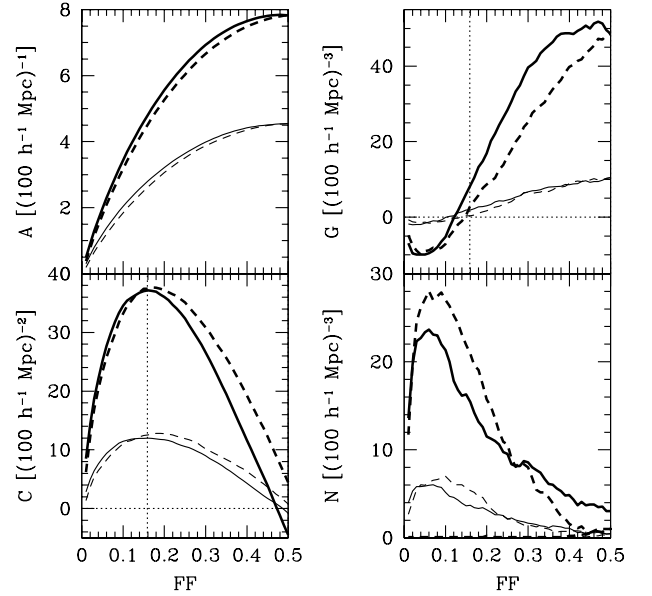} }
\caption{Symmetry test for the density field smoothed with 
  L$_s$=5 h$^{-1}$ Mpc (thick lines) and 10 h$^{-1}$ Mpc (thin lines).
  Global area, curvature and genus at $FF < 0.5$ are shown as a
  function of $1-FF$ by dashed lines.  In addition, the sign of the
  global curvature is also changed.  The number of voids is shown by
  dashed line. The vertical dotted lines mark $FF=0.16$ where the
  Gaussian genus curve crosses the zero level and the integrated mean
  curvature has a maximum.  }
\label{fig:sym_glob}
\end{figure}

In Gaussian fields three global \mf ~(area A, integrated mean curvature
C, and genus G) have certain symmetries as functions of the filling
factor.  In addition, the number of positive peaks, $N_+$ in the excursion
set equals the number of negative troughs, $N_-$:
\ber
A(1-FF_C) &=& A(FF_C),\\
C(1-FF_C) &=& -C(FF_C),\\
G(1-FF_C)&=&G(FF_C),\\
N_-(1-FF_C)&=&N_+(FF_C).  
\eer 
The departure from these symmetries manifests the non-Gaussianity of
the field even before a  comparison with the corresponding Gaussian
functions.  Figure \ref{fig:sym_glob} shows the violation of the
symmetry in the global \mf ~in the \L CDM model smoothed with $L_s$=5
$h^{-1}$Mpc and 10 h$^{-1}$ Mpc. 
The half of every curve corresponding to positive thresholds
($\delta_{\rm TH} \ggeq 0$) is plotted as a solid line and the other half
as a dashed line.
The curves corresponding to smaller
smoothing scale have higher amplitudes and are shown by thick lines.
One can see that while the curves corresponding to $L_s$=5 $h^{-1}$Mpc
are likely to be non-Gaussian,they almost overlap at $L_s$=10
$h^{-1}$Mpc.
\section{Summary and Discussion}
We have studied the large--scale network in the dark matter density field
in real space obtained from N-body simulations by VIRGO consortium
\citep{jenkcdm}.  The density distribution has been smoothed with a
Gaussian window ($L_s=5\ h^{-1}\ Mpc$). 

The major goal of the study was quantitative description and
comparison of over-dense and under-dense regions which were dubbed as
superclusters and voids respectively. We conducted the measurements of
four \mf ~for every individual supercluster and void selected at 99
density thresholds equispaced in terms of the filling factor (Eqs.
\ref{eq:FF-def}, \ref{eq:ffc} and \ref{eq:ffv}) with the SURFGEN code.
By computing the shapefinders (Eqs. \ref{eq:TBL} and
\ref{eq:shapefinder}) from \mf ~we estimated the sizes and shapes of
every supercluster and void. In this study we did not assume any shape
or other characteristics for the large-scale objects.  Instead we have
measured the morphology and various other characteristics of the
objects as they were defined by the isodensity surfaces.  It is worth
stressing that in our study the superclusters and voids have been
defined as three-dimensional regions with closed boundary 
surfaces\footnote{In
some studies voids are viewed as the regions between filaments
connected in a closed loop; for example, see Fig.4 in \cite{she-wey03}.}.

The main results of our investigation are summarized below:
\begin{itemize}
\item Individual superclusters totally occupy no more than about 5\%
  of the total volume and comprise no more than 20\% of mass if the
  largest (\ie percolating) supercluster is excluded (Fig. \ref{fig:isol_vm}).
\item The maximum of the total volume and mass comprised by all 
superclusters except 
  the largest one is reached approximately at
  the percolation threshold: $\delta \approx 1.8$ corresponding to
  $FF_C \approx 0.07$.
\item Individual voids totally occupy no more than 14\% of volume and
  contain no more than 4\% of mass if the largest void is excluded 
 (Fig. \ref{fig:isol_vm}).
\item The maximum of the total volume and mass comprised by all voids
  except the largest one is reached at about the void
  percolation threshold: $\delta \approx -0.5$ corresponding to
  $FF_V \approx 0.22$.
\item Between these two percolation thresholds all superclusters and voids
  except the largest ones take up no more than about 10\% of volume and
  mass. Both largest supercluster and void span throughout the whole
  space and have a very large genus. Therefore they have no well defined 
  sizes, volumes, masses or easily defined shapes.
\item Although superclusters are more massive and voids are more
  voluminous the difference in maximum volumes reached at the corresponding
  percolation thresholds is not much greater than 
  an order of magnitude (see Fig. \ref{fig:mvrho}). The difference in
  maximum masses is even smaller than the difference in volumes.
\item The volumes, masses and geometrical sizes of superclusters
  increase as the density threshold decreases and reach maximum values
  at about the percolation threshold ($\delta \approx 1.8$). 
  At lower thresholds all parameters decrease as the threshold
  continues to decrease.
\item The sizes of voids are significantly larger than those of 
  superclusters even in the density field smoothed with $L_s = 5\ h^{-1}\ Mpc$.
\item The length of a quarter of the most massive superclusters 
 exceeds 50 h$^{-1}$ Mpc. The most voluminous voids are even longer: 
25\% of them are longer than 60 h$^{-1}$ Mpc. The longest 
non-percolating supercluster
is as long as 100 h$^{-1}$ Mpc and the longest non-percolating
void is as long as  200 h$^{-1}$ Mpc.
Both are comparable to the size of the box ($239.5\ h^{-1}\ Mpc$)
and therefore may be affected by the boundaries.
\item The genus value of individual superclusters can be $\sim 5$ while
the genus of individual voids can reach $\sim 55$. This implies significant
amount of substructure in superclusters and voids.
\item Voids have considerably more developed substructure than superclusters.
This is in a general agreement with other studies of voids 
\citep{gro_gel00,pee01}.
\item One of our main results is that voids, as defined through the density
field (read dark matter distribution) can be distinctly non-spherical.
Whether this result carries over to voids in galaxy surveys will
depend upon the nature of the baryon-dark matter biasing and also on
whether the density field is sampled in real or in redshift space.
Since gravitational lensing probes the density field directly, our
results are likely to be of some relevance both for ongoing as well as
future weak lensing surveys of large scale structure.
\item The planarities of both superclusters and voids
are quite low $ P \lleq 0.3$. This implies that the pancake-like structures
in the dark matter density in real space are not typical in the
\L CDM model. We are stressing that this conclusion may be affected by the
size of the smoothing window.
\item The percolation thresholds  as well as some other parameters
depend on the smoothing scale
and for smaller smoothing scales or adaptive filtering windows 
the supercluster percolation threshold must decrease ($FF_C^{perc.} < 0.07$)
and the void percolation threshold increase ($FF_V^{perc.} > 0.22$).
\end{itemize}
\section{Acknowledgments}
VS and SS acknowledge support from the Indo-US collaborative project
DST/NSF/US (NSF-RP087)/2001. SS acknowledges the support from the
Observatory Cote d'Azur Nonlinear Cosmology Program (Nice France) in
summer 2003 when most of the analysis was done.  The trip to Nice was
funded by AAS. JVS is supported by the Senior Research Fellowship of
the Council of Scientific and Industrial Research (CSIR), India. The
simulations studied in this paper were carried out by the Virgo
Supercomputing Consortium using computers based at Computing Centre of
the Max-Planck Society in Garching and at the Edinburgh Parallel
Computing Centre.  The data are publicly available at
{\it www.mpa-garching.mpg.de/NumCos}.

\section{Appendix: Global Minkowski Functionals in a Box with
  Periodic Boundary Conditions}

In computing the \mf~ of individual large-scale objects (superclusters
or voids) we assume that the boundary surface of every object is
closed however it may consist of several disconnected parts.  If the
region is cut by a box boundary it is always closed by the
corresponding part of the box face.  We also assume that the partial
\mf ~are measured for all superclusters and voids at the same set of
density thresholds.  In this case the evaluation of the global
Minkowski functionals with periodic boundary conditions can be done as
follows.
\begin{itemize}
\item {\em Volume}: The total volume is obviously independent of the
  assumption of periodicity $V^{(cp)}=V^{(c)}=\sum v^{(c)}_i$ where
  $v^{(c)}_i$ are the volumes of individual superclusters and
  $V^{(cp)}$ and $V^{(c)}$ denote the global volumes in periodic and
  non-periodic boxes.
\item {\em Area}: Let us denote the total area of the superclusters at
  given filling factor $FF\equiv FF_c$ as $A^{(c)}=\sum a^{(c)}_i$ and
  the total area of voids as $A^{(v)}=\sum a^{(v)}_i$. Then, the sum
  of two becomes
\begin{equation}
A^{(c)} + A^{(v)} = 2 A^{(p)} + A^{(box)}
\end{equation}
where $A^{(p)}$ is the total area of the excursion set in the box with
periodic boundary conditions and $A^{(box)}$ is the total area of the
box boundary (see Fig. \ref{fig:glob_curv} for an illustration).  
For periodic boundary conditions the
global area of the over-dense excursion set obviously equals the
global area of the under-dense excursion set since it is the area of
the common interface surface.  Solving the above equation for
$A^{(p)}$ one obtains
\begin{equation}
A^{(p)} = \frac{A^{(c)} + A^{(v)} - A^{(box)}}{2}.
\end{equation}
\item {\it Integrated mean curvature}: 
The isodensity surfaces is constructed by the triangulation described 
in detail in \citet{sh-s-s-sh03}.
It is convenient for the
  further calculation of the global integrated mean curvatures to
  split the sum into two parts: one is the sum over the inner edges
  and the other is the sum over the edges lying on the box boundary
\begin{equation}
C^{(c)} = \sum_{\mbox{all edges}}c^{(c)}_{j}=\sum_{\mbox{in. edges}}c^{(c)}_{k}
+\sum_{\mbox{b. edges}}c^{(c)}_{i}
\label{eq:Cc_g}
\end{equation}
and 
\begin{equation}
C^{(v)} = \sum_{\mbox{all edges}}c^{(v)}_{j}=\sum_{\mbox{in. edges}}c^{(v)}_{k}
+\sum_{\mbox{b. edges}}c^{(v)}_{i}.
\label{eq:Cv_g}
\end{equation}
For the inner edges the contributions to global mean curvatures of the
superclusters and voids have the same magnitude but opposite signs
\begin{equation}
c^{(c)}_{k}= \frac{1}{2}l_k (\pi-\varphi_k)=-c^{(v)}_{k},
\end{equation}
where $\varphi_k$ is the angle between two surface triangles having
common edge as shown in Fig. \ref{fig:glob_curv}. Please note that in
this equation we use the angle between the triangle planes
while in eq. 6 \citep{sh-s-s-sh03} we used the angle
between the normals to the triangles.  
If the edge lies on the box boundary the relation is less trivial.

Consider a region $R_i$ being cut by the box face into two pieces
$R_{i1}$ and $R_{i2}$ as schematically shown in Fig.
\ref{fig:glob_curv}. The dashed lines show the projections of the
opposite faces of the box. The plane of the figure is assumed to be
orthogonal to the edge lying on the faces of the box and having the
length $l_i$. This assumption is made only for convenience of the
illustration and does not affect the calculation.  We consider the
input to the mean integrated curvature from the edge $l_i$. The code
treats the regions $R_{i1}$ and $R_{i2}$ as two separate regions,
therefore
\begin{equation} 
c^{(c)}_i=\frac{1}{2}[l_i(\pi-\varphi_i) +l_i(\pi-\theta_i)]
=\pi l_i  -\frac{1}{2}l_i(\varphi_i+\theta_i).  
\end{equation}
However, if $R_i$ was treated as one region which corresponded to the
periodic boundary conditions then we would have
\begin{equation} 
c^{(cp)}_i=\frac{1}{2}\pi l_i- \frac{1}{2}l_i(\varphi_i+\theta_i).
\end{equation}
Thus, each edge lying on the face of the box contributes to the global
integrated mean curvature extra $\frac{1}{2}l_i\pi$ compared to the
periodic case.  Similar reasoning yields for the adjacent void
\begin{equation} 
c^{(v)}_i=\frac{1}{2} l_i(\varphi_i +\theta_i)
\end{equation}
and
\begin{equation} 
c^{(vp)}_i=-\frac{1}{2}\pi l_i +\frac{1}{2}l_i(\varphi_i+\theta_i).
\end{equation}
Note that $c^{(cp)}_i = - c^{(vp)}_i$ as it should be.

For the face edges we have
\begin{equation}
c^{(v)}_{i} = -c^{(c)}_{i} +\pi l_i
\end{equation}
Now we have all the terms for equations \ref{eq:Cc_g} and
\ref{eq:Cv_g} and after simple and straightforward calculations obtain
the global mean integrated curvatures for superclusters and voids for
periodic boundary conditions in terms of non-periodic quantities
\begin{equation}
C^{(cp)} = \frac{1}{2}(C^{(c)}-C^{(v)}) = -C^{(vp)}.
\end{equation}
\begin{figure}
  \centering \centerline{ \includegraphics[width=3.5in]{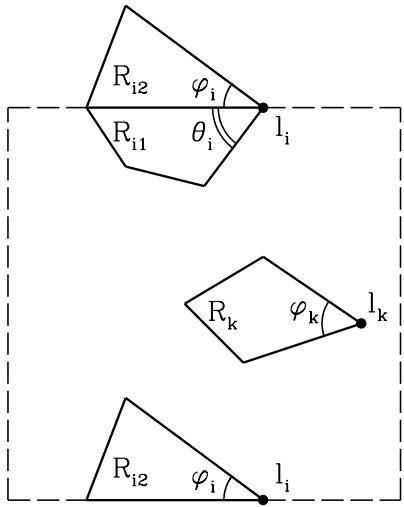} }
\caption{Two superclusters are schematically shown. One is inside the box
(internal) and the 
other is cut into two pieces by the boundary of the box.}
\label{fig:glob_curv}
\end{figure}
\item {\it The Euler characteristic}: Three kinds of elements of 
the triangulated surface are
  triangle faces, edges and vertices. The Euler characteristic of the
  triangulated surface is
\begin{equation}
\chi = F-E+V,
\end{equation}
where $F$, $E$ and $V$ are the total numbers of faces (triangles),
edges and vertices respectively for both closed and open surfaces.

A related measure of topology often used in cosmology is the genus $G
= 1 - \chi/2$.  Although the two quantities are uniquely related and
therefore carry exactly the same information the Euler characteristic
has more convenient mathematical properties.  As we shall see below
including the surfaces with boundaries (\ie open surfaces) into
analysis is very useful in many cases in particular in dealing with
the boundary conditions.  For such surfaces the genus is fractional:
\eg the Euler characteristic of a patch is $\chi_p=1$ and thus its
genus is $G_p=-1/2$. As a result the standard interpretation (eq.
\ref{eq:genus}) cannot be directly applied.  Secondly, and more
importantly, the Euler characteristic is additive while genus is not.
It means that the Euler characteristic of a set of disconnected
surfaces is simply the sum over all members of the set
$\chi_{g}=\sum_i\chi_{i}$ but the genus is not $G_{g} \ne \sum_i g_i$.
For example, three isolated spheres have a genus of $G_g=-2 \ne \sum
g_i=0$ while $\chi_g=\sum\chi_i=6$.  In our study we use Minkowski
functionals for both the global description of the density field and
individual objects. Thus, the additivity of a parameter becomes a very
useful property.

Similarly to the previous discussion of the integrated mean curvature
we split each number into two: one is the input
from the elements lying inside the box and the other from the elements
lying on the boundary \ie faces of the box
\begin{equation}
\chi = (F_{in}-E_{in}+V_{in}) + (F_b-E_b+V_b),
\end{equation}
where subscripts ``in'' and ``b'' correspond to the internal elements of
triangulation lying inside of the box and on the boundaries of the box
respectively.  In the periodic case there are neither triangles nor
edges or vertices on the faces of the box, therefore
\begin{equation}
\chi^{(p)} = F_{in}-E_{in}+V_{in}
\end{equation}
for both superclusters and voids because it is the characteristic of
the common interface surface.

The boundary group of terms gives different results for superclusters
and voids.  In the case of superclusters
$F^{(c)}_b-E^{(c)}_b+V^{(c)}_b = \chi^{(c)}_b$ is the Euler
characteristic of the ``cluster boundary surface'' i.e.  the surface
consisting of all regions on the faces of the box that have been used
for closing the superclusters cut by the boundary.  And in the case of
voids it is obviously the remaining of the box surface which can be
called the ``void boundary surface'' $F^{(v)}_b-E^{(v)}_b+V^{(v)}_b =
\chi^{(v)}_b$.  For instance, consider a simple case when there is
only one cluster cut by a box face. Assuming non-periodic boundary
conditions and closed boundary surfaces for all superclusters the
parts of cluster surface consists of two simply connected pieces each
homeomorphous to a circle and thus $\chi^{(c)}_b= 2$.  The void
boundary surface is homeomorphous to a sphere with two holes (not
tunnels) and thus $\chi^{(v)}_b= 0$.  If there were $n$ superclusters
cut by the box boundaries then $\chi^{(c)}_b= 2n$ and $\chi^{(v)}_b=
2-2n$.  In general, since the parts of the boundary used for closing
the superclusters and the parts used for closing voids make the whole
surface of the box
\begin{equation}
\chi^{(c)}_b + \chi^{(v)}_b= 2 
\end{equation} 
since the box surface is homeomorphous to a sphere and therefore is
equal to two. Summing up the cluster ans void Euler characteristics
one obtains
\begin{equation}
\chi^{(c)} + \chi^{(v)} = 2\chi^{(p)} + \chi^{(c)}_b + \chi^{(v)}_b = 
2(\chi^{(p)}+1).
\end{equation} 
Thus, the global Euler characteristic for periodic boundary
conditions, $\chi^{(p)}$ can be found if both $\chi^{(c)}$ and
$\chi^{(v)}$ are computed at the same threshold under assumption of
non-periodic boundary conditions
\begin{equation}
\chi^{(p)} =\frac{1}{2}(\chi^{(c)} + \chi^{(v)}) -1.
\end{equation}
Computing $\chi^{(c)}$ and $\chi^{(v)}$ also assumes that all
individual clusters and voids have closed surfaces. The surfaces of
superclusters and voids cut by the boundary of the box are closed by
necessary parts of the box boundary surface.
\end{itemize}

\end{document}